\documentclass{emulateapj}

\shorttitle{\lowercase{$r$}-process in the wind from a black-hole
  torus}

\shortauthors{Wanajo and Janka}

\begin{document}

\title{The \lowercase{$r$}-process in the neutrino-driven wind from a
  black-hole torus}

\author{Shinya Wanajo\altaffilmark{1, 2}
        and 
        Hans-Thomas Janka\altaffilmark{2}
        }

\altaffiltext{1}{Technische Universit\"at M\"unchen,
        Excellence Cluster Universe,
        Boltzmannstr. 2, D-85748 Garching, Germany;
        shinya.wanajo@universe-cluster.de}

\altaffiltext{2}{Max-Planck-Institut f\"ur Astrophysik,
        Karl-Schwarzschild-Str. 1, D-85748 Garching, Germany;
        thj@mpa-garching.mpg.de}

\begin{abstract}
  We examine $r$-process nucleosynthesis in the neutrino-driven wind
  from the thick accretion disk (or ``torus'') around a black
  hole. Such systems are expected as remnants of binary neutron star
  or neutron star -- black hole mergers. We consider a simplified,
  analytic, time-dependent evolution model of a $3 M_\odot$ central
  black hole surrounded by a neutrino emitting accretion torus with
  90~km radius, which serves as basis for computing spherically
  symmetric neutrino-driven wind solutions. We find that ejecta with
  modest entropies ($\sim 30$ per nucleon in units of the Boltzmann
  constant) and moderate expansion timescales ($\sim 100$~ms) dominate
  in the mass outflow. The mass-integrated nucleosynthetic abundances
  are in good agreement with the solar system $r$-process abundance
  distribution if a minimal value of the electron fraction at the
  charged-particle freezeout, $Y_\mathrm{e, min} \sim 0.2$, is
  achieved. In the case of $Y_\mathrm{e, min} \sim 0.3$, the
  production of $r$-elements beyond $A \sim 130$ does not reach to the
  third peak but could still be important for an explanation of the
  abundance signatures in $r$-process deficient stars in the early
  Galaxy. The total mass of the ejected $r$-process nuclei is
  estimated to be $\sim 1 \times 10^{-3} M_\odot$. If our model was
  representative, this demands a Galactic event rate of $\sim 2 \times
  10^{-4}$~yr$^{-1}$ for black-hole-torus winds from merger remnants
  to be the dominant source of the $r$-process elements. Our result
  thus suggests that black-hole-torus winds from compact binary
  mergers have the potential to be a major but probably not the
  dominant production site of $r$-process elements.
\end{abstract}


\section{Introduction}

In the past decades, core-collapse supernovae (CCSNe) have been
considered to be the most promising astrophysical site for providing
physical conditions suitable for the $r$(rapid neutron
capture)-process \citep[see, e.g.,][for a recent
review]{Thie2011}. The scenarios include the neutrino-driven wind of
CCSNe \citep{Woos1994, Taka1994, Qian1996, Otsu2000, Wana2001,
  Thom2001, Faro2010}, prompt explosions of CCSNe \citep{Sumi2001} or
of electron-capture supernovae \cite[ECSNe; a subset of CCSNe arising
from collapsing oxygen-neon-magnesium cores,][]{Hill1984, Wana2003},
and the shocked surface layers of the stellar core in collapsing ECSNe
\citep{Ning2007}.

However, recent hydrodynamical simulations of CCSNe
\citep[e.g.,][]{Bura2006, Mare2009} and of ECSNe \citep{Kita2006,
  Jank2008} do not support the prompt explosion or the shocked surface
layer scenarios. The nucleosynthesis calculations with one-dimensional
hydrodynamical results of ECSNe do not confirm the production of
elements beyond $A \sim 90$, either \citep{Hoff2008,
  Wana2009}. Furthermore, recent long-term simulations of CCSNe and
ECSNe show that the neutrino-driven outflows are proton-rich all the
way \citep{Fisc2010, Hued2010}, which poses a severe difficulty to the
scenario that has been favored for a long time, neutrino-driven winds
from proto-neutron stars (PNSs) in CCSNe and ECSNe. Recently,
\citet{Wana2011a} suggested on the basis of two-dimensional models of
ECSNe that a weak $r$-process could lead to the production of
trans-iron elements in the early neutron-rich convective blobs of such
SNe, but no heavier than $A \sim 120$.

In contrast, another popular scenario of the astrophysical
$r$-process, the mergers of double neutron stars \citep[NS-NS,
e.g.,][]{Ross1999, Ruff1999, Shib2000, Shib2006a, Oech2007} or of
black hole -- neutron star binaries \citep[BH-NS, e.g.,][]{Jank1999,
  Shib2006b, Ruff2010}, has not been satisfactorily explored. The
decompression of dynamically ejected neutron-rich crust matter from
NS-NS (or BH-NS) mergers was suggested to be an alternative or
additional $r$-process site \citep{Latt1974, Latt1976, Latt1977,
  Meye1989, Frei1999, Gori2005, Gori2011}. Hyper-massive NSs (HMNSs)
resulting immediately after NS-NS merging \citep[e.g.,][]{Seki2011,
  Rezz2011, Baus2011}, giving rise to magnetically driven and
neutrino-driven outflows for $\sim 10$-100~ms, are also suggested to
eject $r$-processed material \citep{Dess2009}. Furthermore, both NS-NS
(after a possibly only short HMNS phase) and BH-NS (without a HMNS
phase) mergers are expected to form a neutrino radiating accretion
torus around the relic black hole, giving rise to neutrino-driven
winds (hereafter, BH-torus winds) and potential short-duration
gamma-ray-burst (GRB) jets. BH-torus winds are also expected to
provide suitable physical conditions for the $r$-process
\citep{Ruff1999, Surm2008, Metz2008b}.

Merger scenarios have been disfavored compared to those of CCSNe
partly due to discrepancies between Galactic chemical evolution models
and the spectroscopic analyses of Galactic halo stars. The low
Galactic event rate \citep[$7 \times 10^{-6}$--$3 \times
10^{-4}$~yr$^{-1}$,][]{Belc2002} of mergers and the long lifetimes of
binary systems \citep[$\sim 1$~Myr or 100--1000~Myr,][]{Belc2002} are
expected to lead to the delayed appearance of $r$-elements in the
Galactic history with too large star-to-star scattering of their
abundances \citep{Qian2000, Arga2004}. These facts seem to be in
conflict with the observational results of Galactic halo stars
\citep{Hond2004, Fran2007}. \citet{Bane2011} suggested that the early
enrichment of $r$-elements might be due to neutrino-induced
$r$-processing in the compact helium-shells of CCSNe of
low-metallicity stars, and the contribution from mergers could follow
only at a higher metallicity. However, recent studies of Galactic
chemical evolution based on the hierarchical clustering of sub-halos
\citep[][also Y. Ishimaru 2011, in preparation]{Pran2006} or with
various binary population synthesis models \citep{DeDo2004} do not
exclude NS-NS and BH-NS mergers as the dominant astrophysical site of
the $r$-process in the early Galaxy. The reason of observed
star-to-star scattering only found in $r$-elements (but not in
$\alpha$ and iron-group elements) has not been fully understood,
either \citep{Ishi1999, Tsuj1999, Arga2000, Arno2005, Karl2005,
  Cesc2008}. For these reasons, NS-NS and BH-NS mergers cannot be
excluded as the primary source of $r$-elements in the Galaxy. More
studies of nucleosynthesis are highly desired.

In this paper, we examine $r$-process nucleosynthesis in BH-torus
winds, which are expected to be common to both NS-NS and BH-NS
mergers. There exist few previous studies of nucleosynthesis relevant
to these conditions \citep{Surm2008, Metz2009, Caba2011}, which are
based on parametrized outflow conditions. While \citet{Metz2009}
discuss viscously driven mass ejecta, \citet{Surm2008} and
\citet{Caba2011} examined some phenomenologically chosen trajectories
leading to at least a weak $r$-process, but did neither discuss
time-dependences nor did provide the integrated abundance distribution
or the ejected amount of $r$-process nuclei. Currently,
multi-dimensional simulations of the wind phase after the formation of
a stable accretion torus are not yet available. Hence, we adapt a
semi-analytic, spherically symmetric, general relativistic
steady-state wind model for nucleosynthesis calculations of BH-torus
winds. The model has originally been developed for the studies of the
$r$-process in neutrino-driven winds of CCSNe \citep{Wana2001}.

Our paper is organized as follows. In \S~2, we describe our model of
stationary BH-torus winds and discuss basic outcomes derived from the
wind solutions. In \S~3, a phenomenological, time-evolutionary model
of the neutrino luminosities of the torus is introduced, which is
needed to determine the initial composition for nucleosynthesis and to
calculate the mass-integrated yields as well as the ejecta mass. The
results of nucleosynthesis calculations with the wind solutions are
then presented in \S~4, along with the mass-integrated abundances
ejected from the BH torus. In \S~5, we discuss the potential role of
BH-torus winds as the origin of $r$-elements in the Galaxy. A summary
of the paper follows in \S~6.

\section{Modeling BH-torus winds}

\begin{figure}
  \plotone{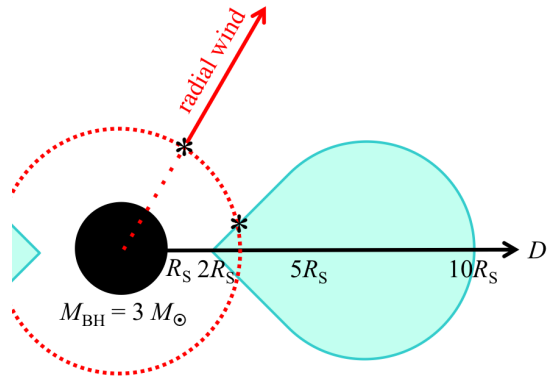}
  \caption{Sketch of our model design for BH-torus winds. A rotating
    BH with the mass $M_\mathrm{BH} = 3 M_\odot$ is located in the
    center of an accretion torus (``neutrino surface'') that lies
    between $2 R_\mathrm{S}$ and $10 R_\mathrm{S}$ from the center,
    where $R_\mathrm{S}$ is the Schwarzschild radius ($=
    8.86$~km). The wind is assumed to be radial. Thus we replace the
    complex neutrino-surface geometry by a spherical outflow model
    considering a neutrinosphere with radius $R_\nu = D$ around the
    gravity center (e.g., the asterisks on the dotted circle).}
\end{figure}

Our treatment of BH-torus winds is based on the semi-analytic,
spherically symmetric, general relativistic model of PNS winds that
has been developed for studies of $r$-process nucleosynthesis by
\citet{Wana2001}. The mass of the central BH is taken to be
$M_\mathrm{BH} = 3 M_\odot$, which can be considered as representative
of NS-NS mergers \citep[or BH-NS mergers with a low-mass
BH,][]{Belc2008}. As illustrated in Figure~1, the rotating accretion
torus around a spinning BH is defined as the ``neutrino surface'' and
is assumed to lie between $2 R_\mathrm{S}$ ($= 17.7$~km) and $10
R_\mathrm{S}$ ($= 88.6$~km) from the center (where $R_\mathrm{S} =
8.86$~km is the Schwarzschild radius) as suggested by detailed
hydrodynamical simulations of NS-NS and BH-NS merging
\citep[e.g.,][]{Jank1999, Ruff1999, Seti2006}\footnote{The radius of
  the inner tip (or innermost stable circular orbit) is smaller for a
  BH co-rotating with the torus than for the non-rotating case. Our
  case, the innermost radius of $D = 2 R_\mathrm{s}$, corresponds to a
  BH spin parameter of $\sim 0.6$ \citep[Fig.1 in][]{Seti2006}. The BH
  rotation is expected not to lead to major differences in the density
  or temperature distributions except for the innermost region
  \citep{Seti2006} but to a moderate increase of the torus mass
  \citep[$\sim 50\%$,][]{Seti2006}. Because the innermost winds make
  an only small contribution to the integrated abundances, and we
  consider the case with $M_\mathrm{ej} \ll M_\mathrm{torus}$ (cf.,
  Eqs.~(11) and (16)), the effects of BH rotation are expected to be
  secondary.}.

In order to connect the aspherical configuration of the BH-torus
system to our spherical model for the wind outflows, an arbitrary
point on the torus is replaced by a point on a hypothetical
neutrinosphere with the same distance from the center, $R_\nu = D$
(dotted circle in Figure~1). The wind trajectory reaching away from
the neutrinosphere, which yields a description of the dynamical and
thermodynamical outflow properties, is then derived in the same manner
as for spherical PNS winds by solving the general relativistic
stationary equations of mass, momentum, and energy conservation
\citep[Eqs.~(1)-(3) in][]{Wana2001}. The Schwarzschild geometry due to
the central BH is included here, but the gravitational effects of the
torus and wind masses are very small and thus neglected. Rotation of
the mass-losing object is neglected as well. The equation of state for
ions (ideal gas) and arbitrarily degenerate, arbitrarily relativistic
electrons and positrons is taken from \cite{Timm2000}. The average
neutrino energies are taken to be 15, 20, and 30~MeV, for electron
neutrino, electron antineutrino, and heavy-lepton neutrinos,
respectively \citep{Jank1999, Seti2006}. The neutrino luminosities of
all the flavors are assumed to have the same value $L_\nu$ in our wind
model. The electron fraction (number of protons per nucleon),
$Y_\mathrm{e}$, is assumed to be 0.5 (for a justification,
see\footnote{As described in \S~3, $L_{\nu_{\mu, \tau}} \ll
  L_{\nu_\mathrm{e}} < L_{\bar{\nu}_\mathrm{e}}$ and $Y_\mathrm{e} \ll
  0.5$ are expected in the early phase of BH-torus outflows. However,
  assuming $L_\nu \equiv \frac{1}{2}(L_{\nu_\mathrm{e}} +
  L_{\bar{\nu}_\mathrm{e}})$ for all neutrino kinds and $Y_\mathrm{e}
  = 0.5$ is reasonably good for computing the dynamics of BH-torus
  winds because the wind-driving energy deposition by $\nu_\mathrm{e}$
  and $\bar{\nu}_\mathrm{e}$ is very similar while that of
  heavy-lepton $\nu$'s is small. Note that these assumptions for
  $L_\nu$ and $Y_\mathrm{e}$ are applied only in this section.}). At
the inner boundary, the density is taken to be $\rho =
10^{10}$~g~cm$^{-3}$ (a different choice has no big
impact\footnote{The wind profile hardly depends on the boundary
  density for $\gtrsim 10^{10}$~g~cm$^{-3}$. Tests show that the wind
  profiles remain essentially unchanged when the inner boundary
  density is varied between values somewhat above $10^9$~g~cm$^{-3}$
  (where the density gradient of the wind changes; Fig.~3, left-middle
  panel) and below the neutrinospheric value (around
  $10^{11}$~g~cm$^{-3}$ or higher).}), and the temperature, $T$, is
taken such that neutrino heating and cooling balance each other ($\sim
$ a few MeV). The velocity (or equivalently, mass ejection rate, $\dot
M$) at the neutrinosphere is determined such that the wind becomes
supersonic through a sonic point.

\begin{figure}
  \plotone{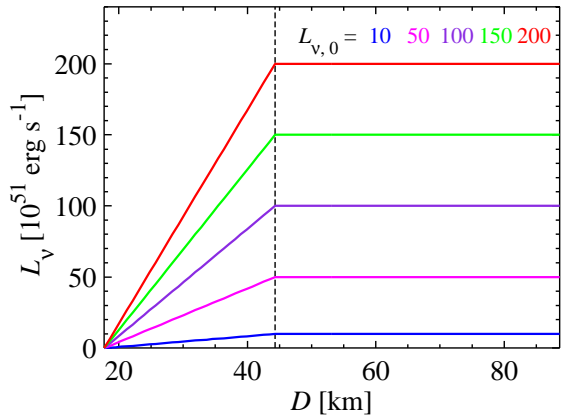}
  \caption{Assumed neutrino luminosities ($L_\nu$) as functions of the
    distance from the center ($D$). $L_\nu$ is assumed to increase
    linearly from $L_{\nu, 0}/100$ to $L_{\nu, 0}$ (denoted in the
    legend in units of $10^{51}$~erg~s$^{-1}$) between $2
    R_\mathrm{S}$ ($=17.7$~km) and $5 R_\mathrm{S}$ ($=44.3$~km;
    dotted line) and adopt a constant value at larger distances $D$.}
\end{figure}

As anticipated from Figure~1, the neutrino flux from the outer regions
of the torus is shielded in the vicinity of the BH by the presence of
the torus itself. In order to mimic this effect in our spherical
models, we simply assume that $L_\nu$ increases linearly from $L_{\nu,
  0}/100$ to $L_{\nu, 0}$ between $2 R_\mathrm{S}$ ($=17.7$~km) and $5
R_\mathrm{S}$ ($=44.3$~km) and adopts a constant value on the outer
side. Figure~2 shows the assumed profiles for selected $L_{\nu, 0}$ in
units of $10^{51}$~erg~s$^{-1}$. We define the outflows from $R_\nu <
5 R_\mathrm{S}$ and $R_\nu > 5 R_\mathrm{S}$ as the ``inner'' and
``outer'' winds, respectively. The calculated radial profiles of
velocity $u$, density $\rho$, and temperature $T$ for the cases with
$R_\nu = 2 R_\mathrm{S}$ (solid lines), $5 R_\mathrm{S}$ (dashed
lines), and $10 R_\mathrm{S}$ (long-dashed lines) are shown in Fig.~3
(left panels) in dependence on the distance $r$ from the
center. $L_{\nu, 0}$ is taken to be $10^{53}$~erg~s$^{-1}$. Right
panels in Fig.~3 display the temporal evolution of $r$, $\rho$, and
$T$ of selected ejected mass elements, where the time coordinate $t =
0$ is set at the inner boundary ($\rho = 10^{10}$~g~cm$^{-3}$).

\begin{figure*}
  \plotone{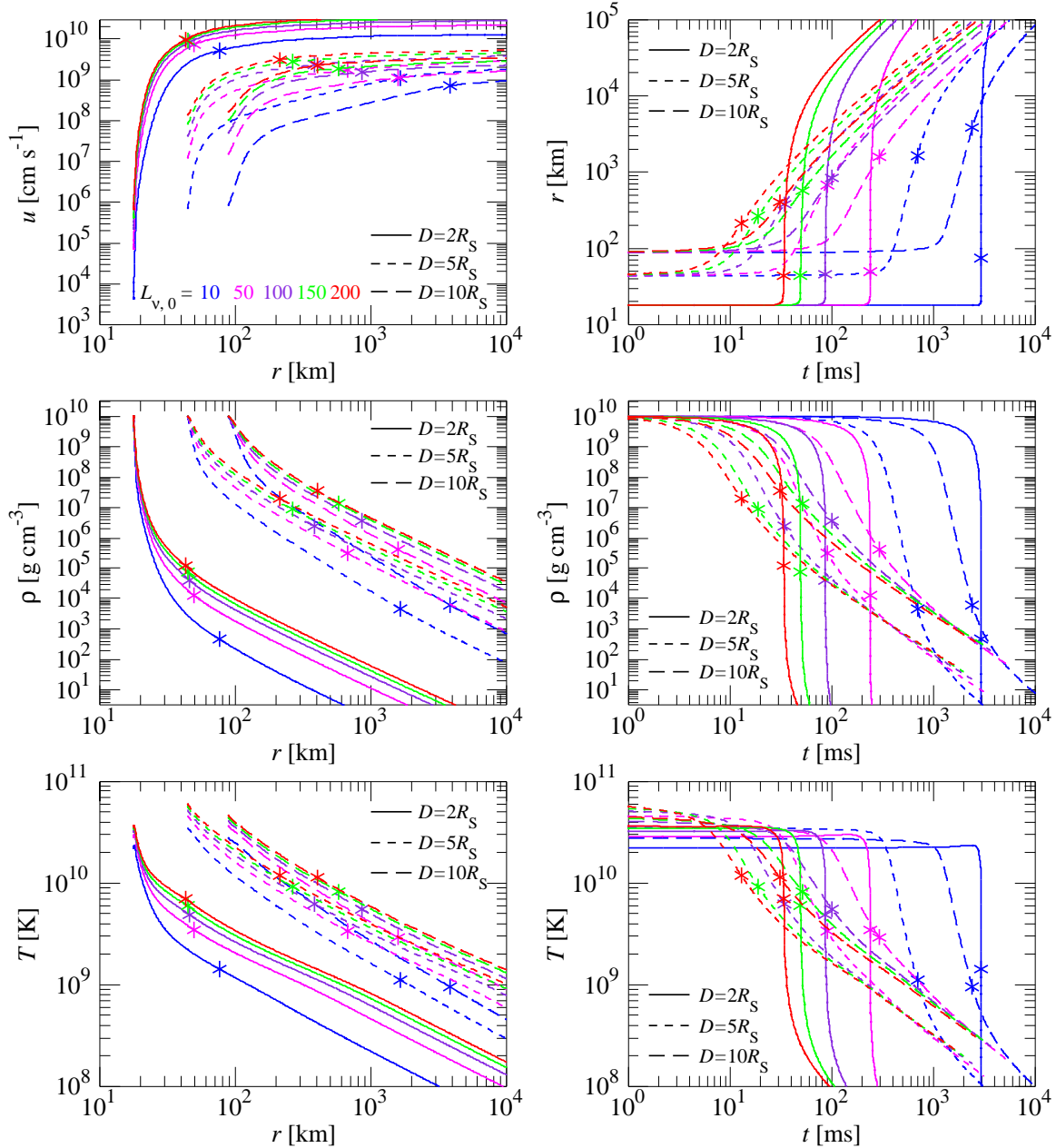}
  \caption{Left: Radial profiles of velocity (top), density (middle),
    and temperature (bottom) for the cases with $R_\nu = 2
    R_\mathrm{S}$ (solid lines), $5 R_\mathrm{S}$ (dashed lines), and
    $10 R_\mathrm{S}$ (long-dashed lines) versus distance $r$ from the
    center. $L_{\nu, 0}$ is taken to be
    $10^{53}$~erg~s$^{-1}$. Asterisks mark the sonic points. Right:
    Temporal evolution of radius (top), density (middle), and
    temperature (bottom) of ejected mass elements for the same
    cases. Time is set to $t = 0$ at the inner boundary ($\rho =
    10^{10}$~g~cm$^{-3}$).}
\end{figure*}

\begin{figure}
  \plotone{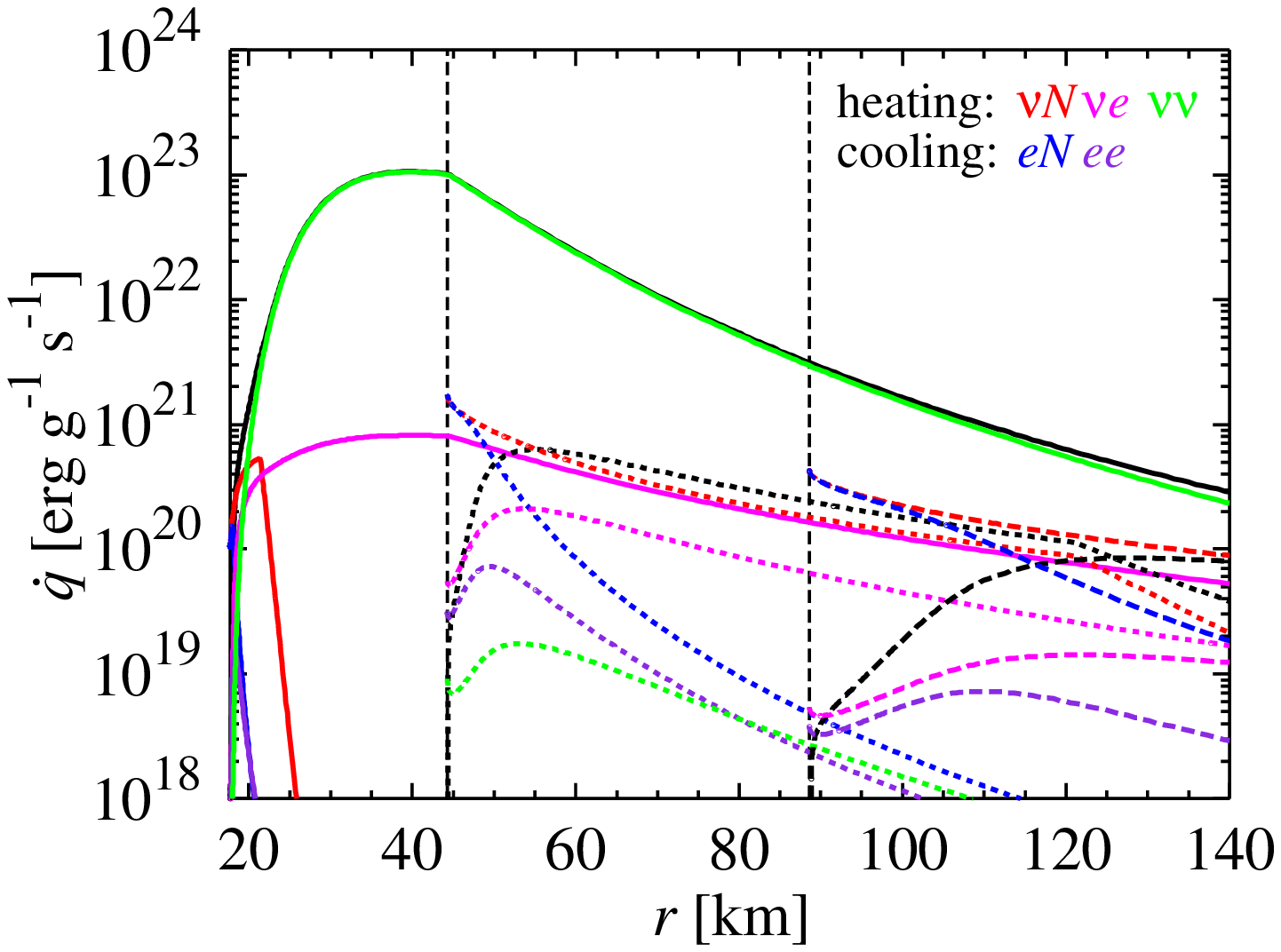}
  \caption{Neutrino heating and cooling rates, $\dot{q}$, for the
    cases with $R_\nu = 2 R_\mathrm{S}$ (solid lines), $5
    R_\mathrm{S}$ (dotted lines), and $10 R_\mathrm{S}$ (dashed lines)
    as functions of radius from the center, $r$. $L_{\nu, 0}$ is taken
    to be $10^{53}$~erg~s$^{-1}$. Heating is due to $\nu_e$ and
    $\bar{\nu}_e$ captures on free nucleons ($\nu N$), neutrino
    scattering by $e^-$ and $e^+$ ($\nu e$), and $\nu \bar{\nu}$ pair
    annihilation to $e^- e^+$ pairs ($\nu \nu$). Cooling is due to
    $e^-$ and $e^+$ captures on free nucleons ($e N$) and $e^- e^+$
    pair annihilation to $\nu \bar{\nu}$ pairs ($ee$). The net total
    rates are indicated in black. The vertical dashed line indicates
    $D = 5 R_\mathrm{S}$. Note that $L_\nu$ increases with $r$ as in
    Fig.~2 for the $R_\nu = 2 R_\mathrm{S}$ case and that the steep
    drop of $\dot{q}_{\nu N}$ for this case is due to the decreasing
    free nucleon abundance by $\alpha$-particle formation.}
\end{figure}

The above assumption of the neutrino luminosity profiles (Fig.~2)
implies fundamental differences in the neutrino heating properties
between the inner and outer winds. In Figure~4, the neutrino heating
and cooling rates, $\dot{q}$ (in units of erg~g$^{-1}$~s$^{-1}$), for
the cases with $R_\nu = 2 R_\mathrm{S}$ (solid lines), $5
R_\mathrm{S}$ (dotted lines), and $10 R_\mathrm{S}$ (dashed lines) are
displayed as functions of the radial distance from the center,
$r$. $L_{\nu, 0}$ is taken to be $10^{53}$~erg~s$^{-1}$. Heating is
due to $\nu_e$ and $\bar{\nu}_e$ captures on free nucleons ($\nu N$),
neutrino scattering by $e^-$ and $e^+$ ($\nu e$), and $\nu \bar{\nu}$
pair annihilation to $e^- e^+$ pairs ($\nu \nu$). Cooling is caused by
$e^-$ and $e^+$ captures on free nucleons ($eN$) and $e^- e^+$ pair
annihilation to $\nu \bar{\nu}$ pairs ($ee$). All the rates are taken
from \citet[][see their Eqs.(8)--(16)]{Otsu2000}, where the
gravitational redshift of neutrino energies and the bending of
trajectories due to general relativistic effects are fully taken into
account.

\begin{figure}
  \plotone{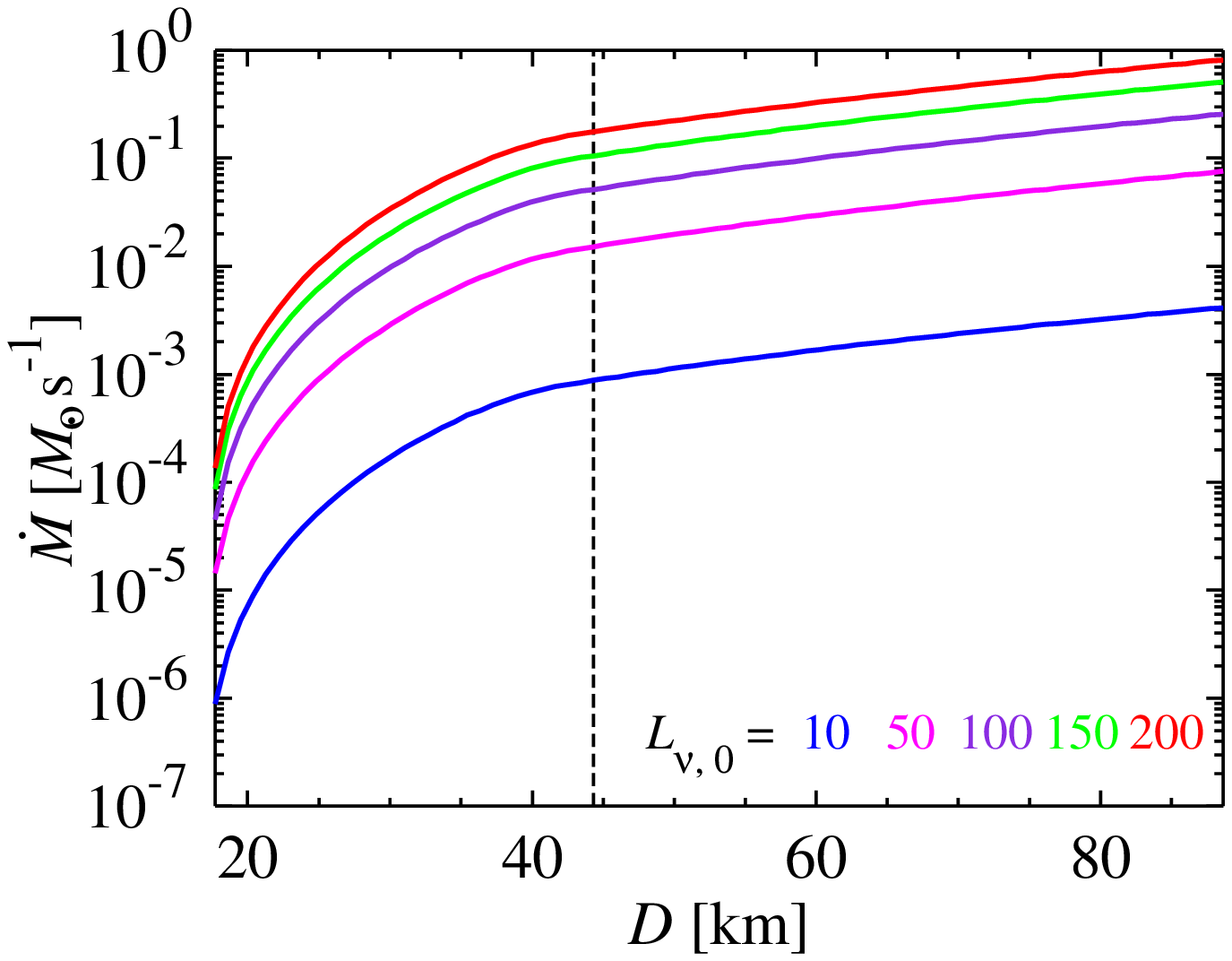}
  \caption{Mass ejection rates, $\dot{M}$, from the equivalent
    neutrinosphere $R_\nu = D$, as a function of $D$ for $L_{\nu, 0}$
    (in units of $10^{51}$~erg~s$^{-1}$) denoted in the legend. The
    vertical dashed line indicates $D = 5 R_\mathrm{S}$.}
\end{figure}

As can be seen in Figure~4, $\dot{q}_{\nu N}$ dominates the heating
rates in the outer winds (dotted and dashed lines). The rate
$\dot{q}_{\nu e}$ plays only a secondary role, and $\dot{q}_{\nu \nu}$
is negligible. This can also be found in the studies of PNS winds
\citep{Qian1996, Otsu2000, Thom2001}. In contrast, in the innermost
winds (solid lines), $\dot{q}_{\nu \nu}$ plays the dominant role for
heating. This is a consequence of the substantially smaller mass
ejection rate, $\dot{M}$ (Fig.~5; defined as that from the
corresponding neutrinosphere), driven by the small $L_\nu$ from the
innermost region of the torus (Fig.~2). This leads to very small
values of $\rho$ and also of $T$ in the vicinity of the torus
\citep[cf.][for a PNS case]{Wana2006b}. For a given radius $r$, the
heating rates scale with $L_\nu$, $\rho$, and $T$ according to
$\dot{q}_{\nu N} \propto L_\nu$, $\dot{q}_{\nu e} \propto L_\nu
\rho^{-1} T^4$, and $\dot{q}_{\nu \nu} \propto {L_\nu}^2
\rho^{-1}$. As a result, the reduction of $\rho$ and $T$ at small $r$
boosts $\dot{q}_{\nu \nu}$ due to radially increasing $L_\nu$ (as
given in Fig.~2) to much higher values than the other rates \citep[see
a similar discussion for anisotropic PNS winds,][]{Wana2006b}. Indeed,
this has been discussed as a promising mechanism to power short GRB
jets presumably arising from NS-NS or BH-NS merging \citep[][and
references therein]{Ruff1999, Jank1999}.

\begin{figure}
  \plotone{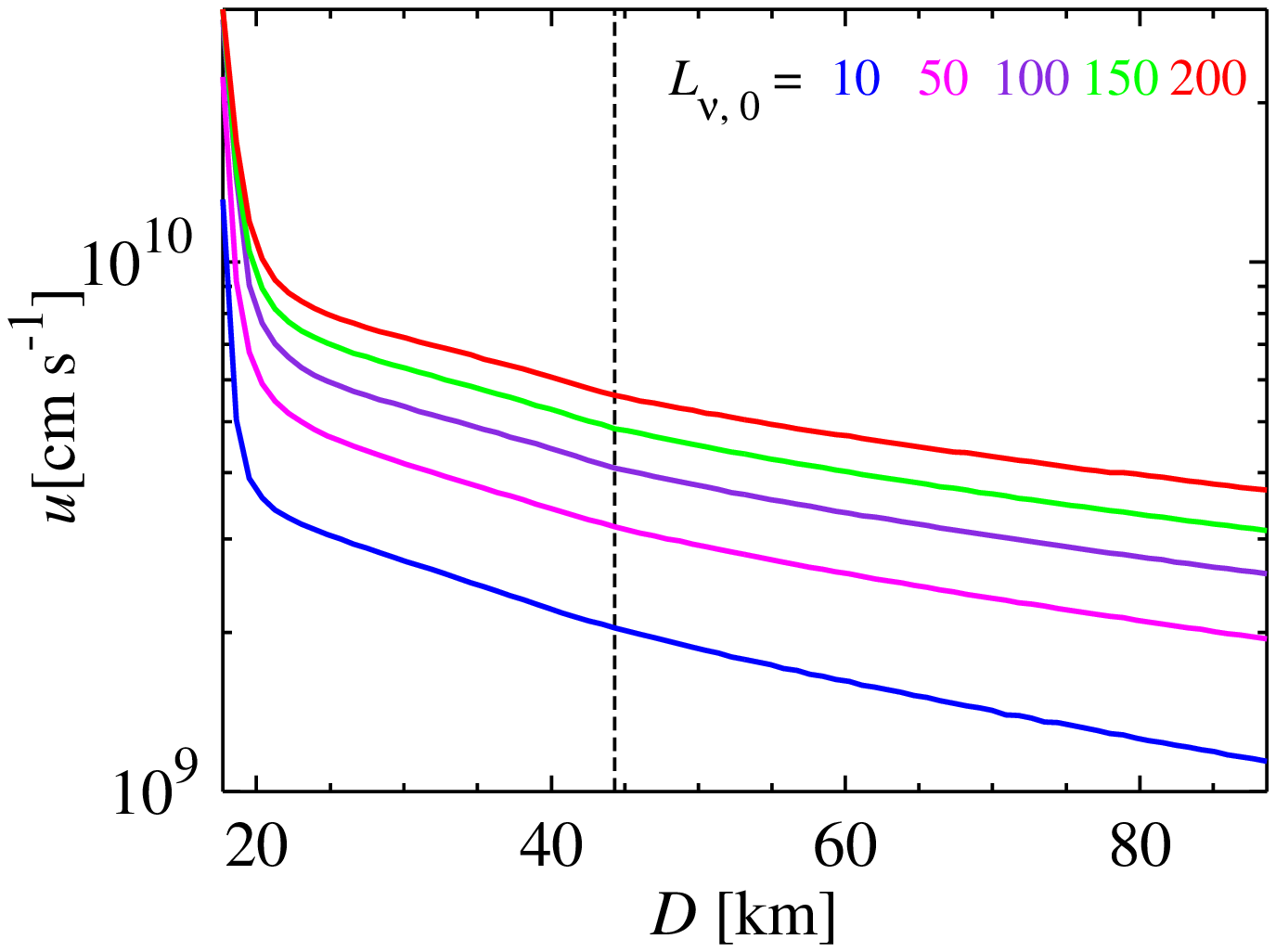}
  \caption{Terminal radial velocities as functions of $D$ for $L_{\nu,
      0}$ (in units of $10^{51}$~erg~s$^{-1}$) denoted in the
    legend. The upper bound is the velocity of light. The vertical
    dashed line indicates $D = 5 R_\mathrm{S}$.}
\end{figure}

\begin{figure}
  \plotone{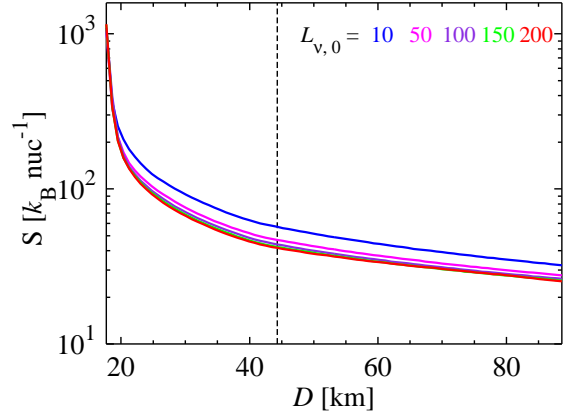}
  \caption{Asymptotic entropies as functions of $D$ for $L_{\nu, 0}$
    (in units of $10^{51}$~erg~s$^{-1}$) denoted in the legend. The
    vertical dashed line indicates $D = 5 R_\mathrm{S}$.}
\end{figure}

\begin{figure}
  \plotone{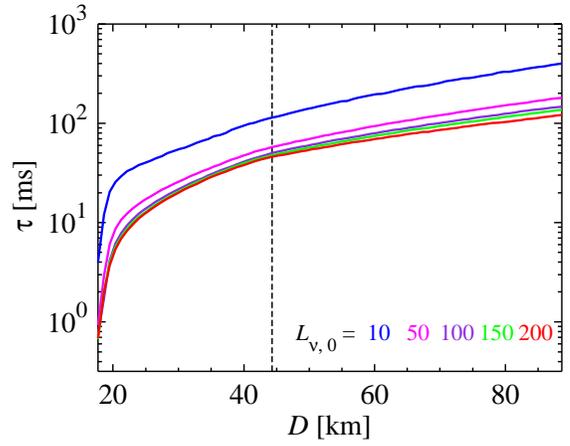}
  \caption{Expansion timescales (the $e$-folding times of temperature
    below $0.5$~MeV) as functions of $D$ for $L_{\nu, 0}$ (in units of
    $10^{51}$~erg~s$^{-1}$) denoted in the legend. The vertical dashed
    line indicates $D = 5 R_\mathrm{S}$.}
\end{figure}

\begin{figure}
  \plotone{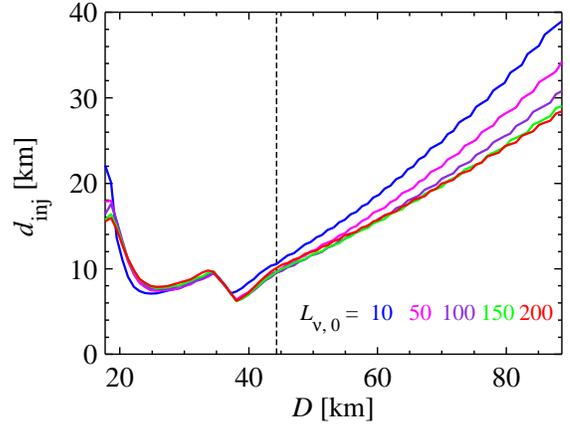}
  \caption{Injection distances, $d_\mathrm{inj}$, as functions of $D$
    for $L_{\nu, 0}$ (in units of $10^{51}$~erg~s$^{-1}$) denoted in
    the legend. Each wind reaches the radius with the maximal
    $\dot{q}$ at $r = D + d_\mathrm{inj}$. The vertical dashed line
    indicates $D = 5 R_\mathrm{S}$.}
\end{figure}

\begin{figure}
  \plotone{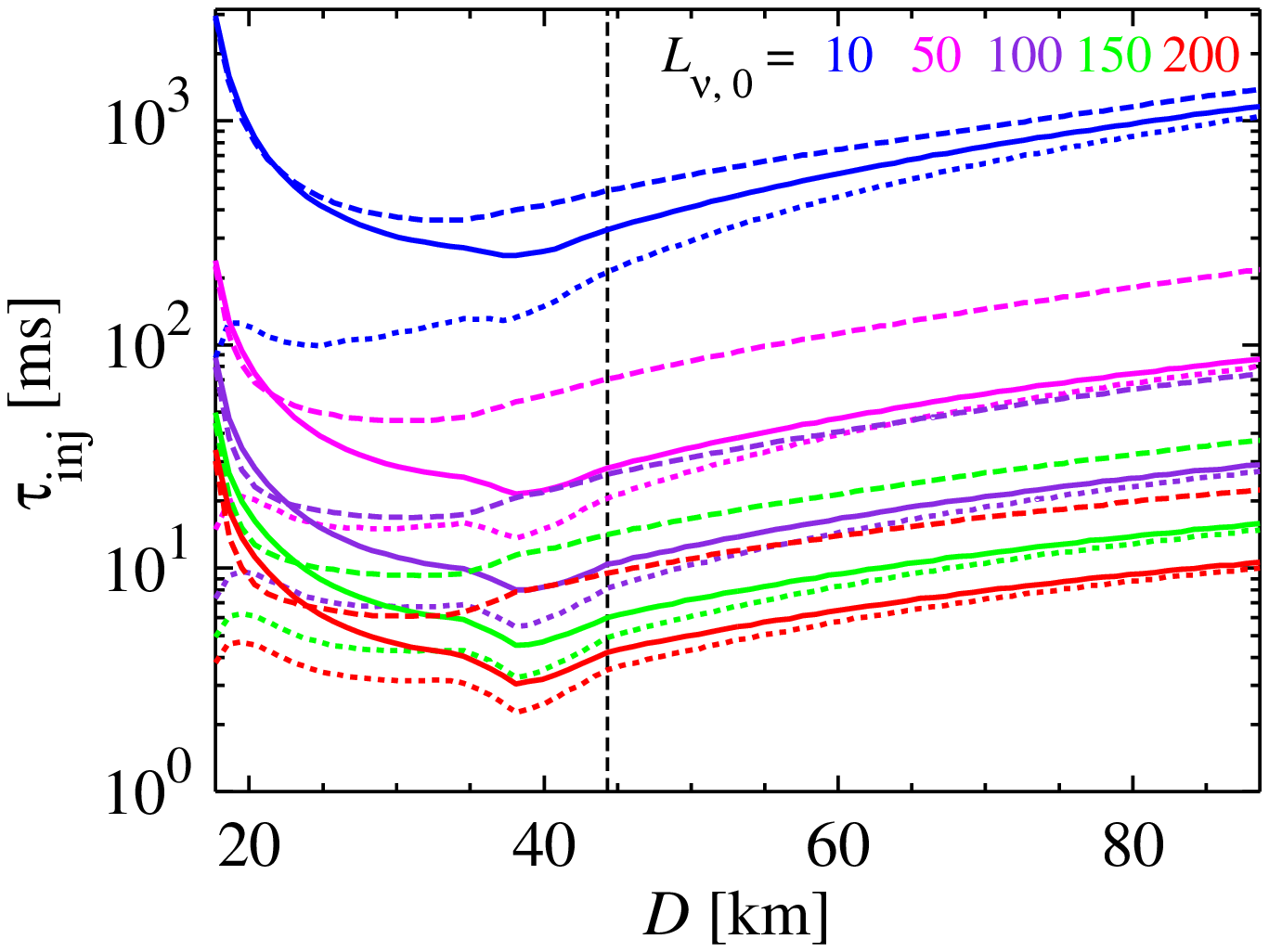}
  \caption{Injection timescales, $\tau_\mathrm{inj}$, as functions of
    $D$ for $L_{\nu, 0}$ (in units of $10^{51}$~erg~s$^{-1}$) denoted
    in the legend. Each wind reaches the radius with the maximal
    $\dot{q}$ at $t =\tau_\mathrm{inj}$, where $t$ is set to 0 at $r =
    D$ (solid lines) or at $r = D + 1$~km (dotted lines; see
    text). Dashed lines indicate the elapsed times from $r = D + 1$~km
    to the sonic points. The vertical dashed line indicates $D = 5
    R_\mathrm{S}$.}
\end{figure}

The greater $\dot{q}_{\nu \nu}$ in the inner winds leads to
substantially faster (terminal) radial velocities, $u$, higher
asymptotic entropies per nucleon, $S$, and short expansion timescales,
$\tau$ (defined as the $e$-folding time of $T$ below 0.5~MeV). As
shown in Figure~6, the innermost winds become relativistic, achieving
$u \sim$ (0.3--1) $c$ (where $c$ is the velocity of light). Figures~7
and 8 also show that the innermost winds obtain extremely high $S$
($\sim 1000$ in units of the Boltzmann constant, $k_\mathrm{B}$) and
short $\tau$ ($\sim 1$~ms). In contrast, the outer winds have modest
$u$ ($\sim 0.1 c$), $S$ ($\sim$ a few 10~$k_\mathrm{B}$) and $\tau$
($\sim$ a few 100~ms). For a given $D$, a larger $L_{\nu, 0}$ leads to
higher $u$ and shorter $\tau$ because of the greater neutrino energy
deposition. However, the resulting larger $\dot{M}$ reduces the
neutrino heating per mass, leading to slightly smaller $S$.

Note that the three-dimensional simulations of BH accretion tori by
\citet{Seti2006} indicate that the matter in the inner region moves
quickly towards the central BH with short accretion timescales of
$\sim 1$--10~ms (with the radial velocities of $\sim (0.01$--$0.1)
c$). Moreover, the lifetime of an accretion torus is estimated to be
of the order of $\sim 100$~ms. This is substantially shorter than the
period of a PNS wind phase ($\sim 10$~s). It is important, therefore,
to test the applicability of our steady-state wind solutions to the
BH-torus case. More specifically, the matter should escape from the
BH-torus surface region within a sufficiently short period of time
($\ll 100$~ms).

To test this, we introduce the ``injection timescale''
$\tau_\mathrm{inj}$ for each wind, during which the matter moves from
the inner boundary to the heating region where $\dot{q}$ peaks
(Fig.~4). Once the matter reaches this maximal-heating region, where
the density is already significantly lower, the subsequent matter
motion would not be substantially disturbed by the time-evolving torus
conditions. We thus regard our steady-state model applicable if
$\tau_\mathrm{inj} \ll 100$~ms is satisfied. Figure~9 shows the
injection distances, defined as $d_\mathrm{inj} = r(\dot{q} =
\dot{q}_\mathrm{max}) - D$, as functions of $D$ for selected $L_{\nu,
  0}$. The inner (except for innermost) winds have relatively small
$d_\mathrm{inj}$ ($\sim 10$~km), while the outer winds have greater
$d_\mathrm{inj}$ increasing with $D$ (from 10~km to 30--40~km). Note
that $d_\mathrm{inj}$ is relatively insensitive to $L_{\nu, 0}$ as
also found in the PNS case \cite[see Fig.~4 in][]{Thom2001}. Figure~10
displays the injection timescales $\tau_\mathrm{inj}$ (solid lines) as
functions of $D$ for selected $L_{\nu, 0}$. We find particularly large
$\tau_\mathrm{inj}$ for the innermost winds (see also right panels in
Fig.~3). It should be noted, however, that the injection timescales
could be overestimated owing to the ambiguity of the boundary density
(see footnote 5) and thus of the boundary velocity (cf., $\dot{M} =
4\pi r^2 \rho u$). To avoid this, the elapsed times from $r = D +
1$~km to $r = D + d_\mathrm{inj}$ are also drawn in Fig.~ 10 (dotted
lines). This indicates that the large $\tau_\mathrm{inj}$ for the
innermost winds is largely due to the small initial velocities (see
Fig.~3, top-left panel) arising from our choice of the boundary
density ($\rho = 10^{10}$~g~cm$^{-3}$ for all the cases).

Considering the condition $\tau_\mathrm{inj} \ll 100$~ms, we find that
our steady-state treatment can be (at least marginally) applicable for
the winds with $L_{\nu, 0} > 50$ (in units of
$10^{51}$~erg~s$^{-1}$). Note that, once the matter reaches $r
=r(\dot{q}_\mathrm{max}$), the outflow becomes supersonic within a
short period of time. The supersonic outflows are sonically
disconnected from the surface and are not affected by the
time-evolving BH-torus any more. Dashed lines in Fig.~10 indicate the
elapsed times from $r = D + 1$~km to the sonic points, which are $\sim
2 \tau_\mathrm{inj}$ except for the innermost region. The total ejecta
mass might be somewhat overestimated owing to the longer
$\tau_\mathrm{inj}$ for lower $L_{\nu, 0}$. As we find in Fig.~5,
however, $\dot{M}$ for $L_{\nu, 0} < 50$ (in units of
$10^{51}$~erg~s$^{-1}$) is substantially smaller and its contribution
to the total mass will be sub-dominant. The ejecta mass from the inner
region might be also overestimated because of the fast accretion
timescales estimated from hydrodynamical studies \citep[$\sim
1$--10~ms,][]{Seti2006}. As we will see below, however, the ejecta
from the outer region dominate the total amount ($\sim 90\%$;
\S~3). It is also important to note, taking our injection timescale
condition seriously, that the outer-torus region beyond $\sim 100$~km
is not expected to contribute significantly to the total ejecta mass
because of long injection timescales.

It should be noted that the work by \citet{Seti2006} was based on a
neutrino trapping/leakage scheme and employed a simple $\alpha$-model
for the gas viscosity. The dynamics of outflows from BH tori will be
highly dependent on neutrino transport as well as magnetic fields,
which are to be explored in future studies of hyperaccreting BHs. This
did not allow \citet{Seti2006} to describe mass loss from the tori by
energy transfer and deposition associated with neutrino transport, nor
did their models adequately account for magnetically-driven mass
loss. Outflows and winds were previously discussed by \citet{Prue2006,
  Metz2008a, Metz2008b, Metz2009, Surm2008, Caba2011, Ross2011},
although all of those works did not employ self-consistent
hydrodynamical models of this phenomenon.

\section{Time Evolution of the BH-torus}

\subsection{Why is the BH-torus wind neutron-rich?}

As mentioned in \S~1, recent hydrodynamical simulations of CCSNe
predict that the PNS wind is proton-rich \citep{Fisc2010,
  Hued2010}. The BH-torus wind from ``collapsars'' \citep{MacF1999}
resulting from collapsing rapidly rotating stellar cores is also
suggested to be proton-rich \citep[e.g.,][]{Kizi2010}. The BH-torus
from NS-NS and BH-NS mergers, originating from decompressed, very
neutron-rich NS matter, is different from the collapsar case. The
initially neutron-rich torus leads to more production of
$\bar{\nu}_\mathrm{e}$ by $e^+$ captures on neutrons,
\begin{eqnarray}
n + e^+ \longrightarrow \bar{\nu}_\mathrm{e} + p,
\end{eqnarray}
than $\nu_\mathrm{e}$ production by $e^-$ captures on protons,
\begin{eqnarray}
p + e^- \longrightarrow \nu_\mathrm{e} + n,
\end{eqnarray}
and thus one obtains $L_{\bar{\nu}_\mathrm{e}} >
L_{\nu_\mathrm{e}}$. This allows the BH-torus wind (outside of the
torus at lower densities) to remain neutron-rich owing to more
$\bar{\nu}_\mathrm{e}$ captures on free protons,
\begin{eqnarray}
\bar{\nu}_\mathrm{e} + p \longrightarrow n + e^+,
\end{eqnarray}
than $\nu_\mathrm{e}$ captures on free neutrons,
\begin{eqnarray}
\nu_\mathrm{e} + n \longrightarrow p + e^-.
\end{eqnarray}
However, despite the fact that the neutrinospheric layers in a PNS are
also neutron-rich, the simulations cited above show that PNSs produce
essentially equal luminosities of $\nu_\mathrm{e}$ and
$\bar{\nu}_\mathrm{e}$, resulting in the proton-richness of the ejecta
because of the neutron-proton mass difference, which reduces
$\bar{\nu}_\mathrm{e}$ captures compared to $\nu_\mathrm{e}$
captures. So what exactly is the difference between the PNS and merger
BH-torus cases?  This can be explained as follows.

For PNSs, the main point is the fact that the PNS is in a phase of
\textit{neutronization}. It evolves from more proton-rich (symmetric)
initial conditions to more neutron-rich final conditions. In course of
this transition, its core radiates more $\nu_\mathrm{e}$ than
$\bar{\nu}_\mathrm{e}$ (see Eqs.~(1) and (2)). On their radial way out
of the PNS through highly neutrino-opaque layers this initial
$\nu_\mathrm{e}$ flux is converted into equal $\nu_\mathrm{e}$ and
$\bar{\nu}_\mathrm{e}$ fluxes by multiple absorption and reemission
processes (and $\mu$ and $\tau$ fluxes of similar size). The escaping
$\nu_\mathrm{e}$ and $\bar{\nu}_\mathrm{e}$ luminosities are roughly
equal because the neutrinospheric layers with their typical densities
($\sim 10^{11}$~g~cm$^{-3}$ initially and up to $\sim
10^{13}$~g~cm$^{-3}$ at very late times) reach their so-called
``$\beta$-equilibrium" state (defined by equal number production of
$\nu_\mathrm{e}$ and $\bar{\nu}_\mathrm{e}$ and $Y_\mathrm{e} =$
const. in time) within some tens of milliseconds up to about $\sim
100$~ms. This is much shorter than the PNS cooling timescale of
seconds. On the long PNS evolution timescale the emission is therefore
mostly characterized by essentially equal $L_{\nu_\mathrm{e}}$ and
$L_{\bar{\nu}_\mathrm{e}}$ (with a small difference making sure that
there is a net $\nu_\mathrm{e}$-flux out of the still neutronizing
high-density PNS core).

In contrast, the hot post-merger torus is composed of decompressed
neutron star matter. It starts out from a very neutron rich initial
state and (on its way to a new $\beta$-equilibrium condition)
\textit{protonizes} gradually (Eq.~(1)) because of overall lower
densities than it had initially. Since the density and temperature of
the torus (on average some $10^{11}$--$10^{12}$~g~cm$^{-3}$ and about
3--10~MeV) are very similar to the neutrinospheric conditions of a
PNS, the protonization of the torus proceeds on the same timescale as
the neutronization of the neutrinospheric region in the PNS case. As
mentioned above, this typical timescale is tens of milliseconds at the
given densities, presumably up to $\sim 100$~ms. This is exactly a
major fraction of the accretion timescale of most of the torus mass
into the BH (and is the timescale for the protonization assumed in our
simple model described in \S~3.2). Because of the protonization,
$L_{\bar{\nu}_\mathrm{e}}$ is higher than $L_{\nu_\mathrm{e}}$, in
inverse analogy to the fact that $L_{\nu_\mathrm{e}}$ is higher in the
case of a collapsing stellar core during the early times when the
postshock layer is still on its way to reach the $\beta$-equilibrium
condition, starting out from a nearly symmetric initial state. For
this reason, the neutrino-driven ejecta from the BH-torus are expected
to be neutron-rich (see Eqs.~(3) and (5)) for a significant duration
of the wind phase.

Note that a direct consequence of the fact that the torus conditions
in density and temperature are roughly similar to the neutrinospheric
conditions in a PNS is the finding that the $\mu$ and $\tau$ neutrino
luminosities are 5--10 times lower than the $\nu_\mathrm{e}$ and
$\bar{\nu}_\mathrm{e}$ luminosities in the case of the tori
\citep[see][]{Jank1999, Ruff1999}. In a PNS the $\mu$ and $\tau$
neutrinos are produced at much higher densities than $\nu_\mathrm{e}$
and $\bar{\nu}_\mathrm{e}$. Such densities are not reached by the
accretion tori.

\subsection{Time Evolution of Neutrino Luminosities}

In order to calculate the nucleosynthesis for each wind trajectory,
the $Y_\mathrm{e}$ value of the torus, which defines the initial
composition, should be specified. In addition, the ejecta mass for
each wind trajectory should be determined to calculate the
mass-integrated nucleosynthetic abundances. For this purpose, we
assume that the time evolution of $L_{\nu_\mathrm{e}}$ and
$L_{\bar{\nu}_\mathrm{e}}$ can be written as
\begin{eqnarray}
L_{\nu_\mathrm{e}} (t) & = & L_{\nu, \mathrm{i}} \left(\frac{t}{t_\mathrm{i}}\right)^{-\beta},\\
L_{\bar{\nu}_\mathrm{e}} (t) & = & L_{\nu, \mathrm{i}} \left(\frac{t}{t_\mathrm{i}}\right)^{-\beta}
\left[1+2\left(\frac{t}{t_\mathrm{i}}\right)^{-\alpha}\right],
\end{eqnarray}
where $L_{\nu, \mathrm{i}} = 10^{53}$~erg~s$^{-1}$ is chosen as a
representative value of the initial neutrino luminosity at $t =
t_\mathrm{i} = 10$~ms \citep[e.g.,][]{Jank1999, Seti2006}. The power
exponent $\beta$ is taken to be 1.3 \citep{Metz2008b, Lee2009}. The
value of $\alpha$ is uncertain and for simplicity assumed to be a
modest number, $\alpha = 1$, because $L_{\bar{\nu}_e} > L_{\nu_e}$ is
expected during a significant time of the torus evolution
(\S~3.1). These relations give
$L_{\bar{\nu}_\mathrm{e}}/L_{\nu_\mathrm{e}} = 3$ at $t =
t_\mathrm{i}$, being in agreement with hydrodynamical results
\citep{Jank1999, Seti2006}, and $L_{\bar{\nu}_\mathrm{e}} =
L_{\nu_\mathrm{e}}$ for $t \gg t_\mathrm{i}$
\citep{Metz2008b}. Eqs.~(5) and (6) yield an approximative time
evolution of (the torus-averaged value of) $Y_\mathrm{e}$ in the
protonizing torus (\S~3.1; Eqs.~(1) and (2)),
\begin{eqnarray}
\dot{Y_\mathrm{e}}
= \frac{L_{\bar{\nu}_\mathrm{e}} (t) - L_{\nu_\mathrm{e}} (t)}
        {N_\mathrm{b} \langle \epsilon_\nu \rangle}
= \frac{2 L_{\nu, \mathrm{i}}}{N_\mathrm{b} \langle \epsilon_\nu \rangle}
  \left(\frac{t}{t_\mathrm{i}}\right)^{-(\alpha + \beta)},
\end{eqnarray}
where $N_\mathrm{b}$ and $\langle \epsilon_\nu \rangle$ are the total
baryon number of the torus and the mean neutrino energy \citep[$\sim
18$~MeV,][]{Jank1999, Seti2006} averaged for $\nu_\mathrm{e}$ and
$\bar{\nu}_\mathrm{e}$.  This can be integrated to yield
\begin{eqnarray}
Y_\mathrm{e}(t)
& = & Y_\mathrm{e, i}
      + \frac{2 L_{\nu, \mathrm{i}}}{N_\mathrm{b} \langle \epsilon_\nu \rangle}
      \int_{t_\mathrm{i}}^t \left(\frac{t'}{t_\mathrm{i}}\right)^{-(\alpha + \beta)} dt'\nonumber\\
& = & Y_\mathrm{e, i}
      + \frac{2 L_{\nu, \mathrm{i}}\, t_\mathrm{i}}{(\alpha + \beta -1) N_\mathrm{b} \langle \epsilon_\nu \rangle}
      \left[1-\left(\frac{t}{t_\mathrm{i}}\right)^{-(\alpha + \beta - 1)}\right],
\end{eqnarray}
where $Y_\mathrm{e, i} = 0.05$ is assumed for the initial electron
fraction \citep[e.g.,][]{Ruff1999}. Further assuming that
$Y_\mathrm{e}$ asymptotes to $Y_\mathrm{e, f} = 0.50$ for $t \gg
t_\mathrm{i}$ \citep[e.g.,][]{Metz2008b}, we get
\begin{eqnarray}
\frac{2 L_{\nu, \mathrm{i}}\, t_\mathrm{i}}{(\alpha + \beta -1) N_\mathrm{b} \langle \epsilon_\nu \rangle}
= Y_\mathrm{e, f} - Y_\mathrm{e, i}.
\end{eqnarray}
This simplifies Eq.~(8) to become
\begin{eqnarray}
Y_\mathrm{e}(t)
= Y_\mathrm{e, f}
- (Y_\mathrm{e, f} - Y_\mathrm{e, i})
  \left(\frac{t}{t_\mathrm{i}}\right)^{-(\alpha + \beta - 1)}.
\end{eqnarray}
Eq.~(10) provides $Y_\mathrm{e}$ of the torus at a given evolutionary
time, $t$. Note that the value of $N_\mathrm{b}$ (not made use of in
this study) can be estimated from Eq.~(9) and leads to the approximate
torus mass,
\begin{eqnarray}
M_\mathrm{torus}
= N_\mathrm{b} m_\mathrm{u}
\approx \frac{2 L_{\nu, \mathrm{i}}\, t_\mathrm{i}\, m_\mathrm{u}}{(\alpha + \beta -1) (Y_\mathrm{e, f} - Y_\mathrm{e, i}) \langle \epsilon_\nu \rangle}
\sim 0.1 M_\odot,
\end{eqnarray}
which is consistent with hydrodynamical results of compact binary
mergers \citep[e.g.,][]{Jank1999, Ruff1999}.

\subsection{Connecting Steady Wind and BH-Torus Evolution Models}

Our description of BH-torus winds in \S~2 is based on steady-state
solutions for constant $L_{\nu, 0}$ and $Y_\mathrm{e}$. In order to
connect the computed wind trajectories to the time-evolving torus
model described in \S~3.2, we assume
\begin{eqnarray}
L_{\nu, 0}
=
\frac{L_{\nu_\mathrm{e}}(t) + L_{\bar{\nu}_\mathrm{e}}(t)}{2}.
\end{eqnarray}
This is a reasonable assumption because the wind from the outer torus
is driven predominantly by $\nu_\mathrm{e}$ and $\bar{\nu}_\mathrm{e}$
captures on free nucleons (Fig.~4), whose heating effect is nearly
symmetric between $\nu_\mathrm{e}$ and $\bar{\nu}_\mathrm{e}$. This is
not true for the innermost torus, which however does not significantly
contribute to the mass-integrated nucleosynthetic abundances as we
will see later. Eq.~(12) gives, with Eqs.~(5), (6) and (10),
$L_{\nu_\mathrm{e}}$, $L_{\bar{\nu}_\mathrm{e}}$, $t$, $\Delta t$, and
$Y_\mathrm{e}$ (as the initial value for nucleosynthesis) for a given
$L_{\nu, 0}$ in our steady-state model of BH-torus winds. These values
are listed in Table~1 for $L_{\nu, 0} = 200, 190, \cdots, 10$ (in
units of $10^{51}$~erg~s$^{-1}$). Here, $\Delta t$ are the time
intervals $\Delta t_j = t_{j+1} - t_j$ between the $j$-th and
$(j+1)$-th wind trajectories. For the 20th wind trajectory (last line
in Table~1), $\Delta t$ is determined such that the total evolutionary
time becomes $t = 100$~ms.

\begin{deluxetable}{cccccccc}
\tablecaption{Time-Evolution of the BH Torus}
\tablewidth{0pt}
\tablehead{
\colhead{$j$} &
\colhead{$L_{\nu, 0}$\tablenotemark{a}} &
\colhead{$L_{\nu_\mathrm{e}}$\tablenotemark{a}} &
\colhead{$L_{\bar{\nu}_\mathrm{e}}$\tablenotemark{a}} &
\colhead{$t$ [ms]} &
\colhead{$\Delta t$ [ms]} &
\colhead{$Y_\mathrm{e}$\tablenotemark{b}}
}
\startdata
 1 & 200 & 100   & 300   & 10.0  & 0.290 & 0.050 \\
 2 & 190 &  96.4 & 284   & 10.3  & 0.316 & 0.066 \\
 3 & 180 &  92.6 & 267   & 10.6  & 0.346 & 0.083 \\
 4 & 170 &  88.9 & 251   & 11.0  & 0.381 & 0.100 \\
 5 & 160 &  85.0 & 235   & 11.3  & 0.422 & 0.118 \\
 6 & 150 &  81.0 & 219   & 11.8  & 0.471 & 0.135 \\
 7 & 140 &  77.0 & 203   & 12.2  & 0.530 & 0.153 \\
 8 & 130 &  72.9 & 187   & 12.8  & 0.603 & 0.172 \\
 9 & 120 &  68.6 & 171   & 13.4  & 0.692 & 0.191 \\
10 & 110 &  64.3 & 156   & 14.1  & 0.806 & 0.211 \\
11 & 100 &  59.8 & 140   & 14.9  & 0.953 & 0.231 \\
12 &  90 &  55.1 & 125   & 15.8  & 1.15 & 0.252 \\
13 &  80 &  50.3 & 110   & 17.0  & 1.42 & 0.274 \\
14 &  70 &  45.3 &  94.7 &  18.4 & 1.81 & 0.296 \\
15 &  60 &  40.1 &  79.9 &  20.2 & 2.41 & 0.319 \\
16 &  50 &  34.7 &  65.3 &  22.6 & 3.40 & 0.344 \\
17 &  40 &  28.9 &  51.1 &  26.0 & 5.26 & 0.370 \\
18 &  30 &  22.7 &  37.3 &  31.3 & 9.56 & 0.398 \\
19 &  20 &  16.1 &  23.9 &  40.8 & 24.7 & 0.428 \\
20 &  10 &   8.68 & 11.3 &  65.6 & 34.4 & 0.461 
\enddata
\tablenotetext{a}{in units of $10^{51}$~erg~s$^{-1}$.}
\tablenotetext{b}{initial value for nucleosynthesis.}
\end{deluxetable}

The mass ejection rate for the $j$-th value of $L_{\nu, 0}$ from the
torus region $D_i \le D \le D_{i+1}$ is calculated as
\begin{eqnarray}
\Delta \dot{m}_{i, j} 
= \frac{\dot{M}_{i, j}}{4 \pi D_i^2} 2 \pi (D_{i+1}^2 - D_i^2)
= \frac{\dot{M}_{i, j} (D_{i+1}^2 - D_i^2)}{2 D_i^2},
\end{eqnarray}
where the torus is imagined to be a disk, ejecting matter
perpendicularly toward the north and south directions. Here,
$\dot{M}_{i, j}$ is the mass ejection rate from the corresponding
neutrinosphere with $R_\nu = D_i$ (Fig.~5) for the $j$-th value of
$L_{\nu, 0}$, and the mass flux density $(\dot{M}_{i, j}/4\pi D_i^2)$
of the spherical wind calculation is weighted by the effective torus
surface element in Eq.~(13). The mass ejection rate during time
interval $\Delta t_j$ from the entire torus with $L_{\nu, 0} =
(L_{\nu, 0})_j$ is then given as
\begin{eqnarray}
\dot{m}_j
= \sum_i \Delta \dot{m}_{i, j}.
\end{eqnarray}
Eq.~(13) with the $\Delta t$ from Table~1 gives the ejecta mass
\begin{eqnarray}
\Delta m_{i, j} = \Delta \dot{m}_{i, j} \Delta t_j
\end{eqnarray}
for the wind at $D = D_i$ and with $L_{\nu, 0} = (L_{\nu, 0})_j$. From
Eq.~(15) the total ejecta mass in the BH-torus outflow during the
first 100~ms is calculated as
\begin{eqnarray}
m_\mathrm{ej}
= \sum_{i, j} \Delta m_{i, j}
= 1.96 \times 10^{-3} M_\odot,
\end{eqnarray}
where the inner region ($D \le 5 R_\mathrm{S}$) contributes only $\sim
10\%$ ($= 2.40 \times 10^{-4} M_\odot$). The total ejecta mass in
Eq.~(16) is only a few percent of our assumed (initial) BH-torus mass
($\sim 0.1 M_\odot$; Eq.~(11)). This may justify our assumption of
steady-state neutrino-driven outflows, provided that the lifetime of
the BH-torus is $> 100$~ms, which can be expected to be the case,
considering typical mass-accretion rates of BH-tori in hydrodynamical
models \citep[see, e.g., Fig.~4 in][]{Seti2006}. Note that the initial
high mass-loss phase (with $\dot{M} \gtrsim 0.1 M_\odot$~s$^{-1}$;
Fig.~5) for the outer winds lasts for only a short period of time (a
few ms; Table~1).

\begin{figure}
  \plotone{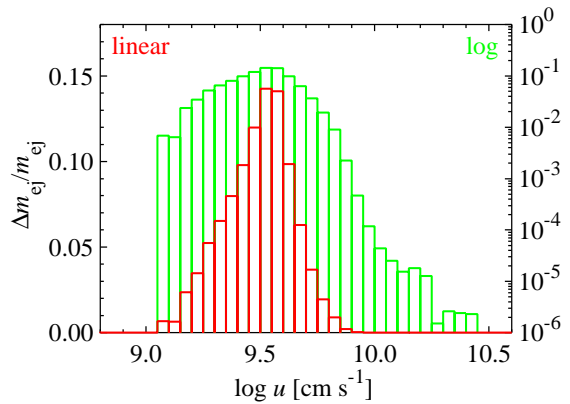}
  \caption{Histogram of the ejecta mass distribution $\Delta
    m_\mathrm{ej}$ (normalized by the total ejecta mass,
    $m_\mathrm{ej}$) in linear (left vertical axis) and logarithmic
    (right vertical axis) scales as a function of the terminal
    expansion velocity $u$ (in logarithmic scale).}
\end{figure}

\begin{figure}
  \plotone{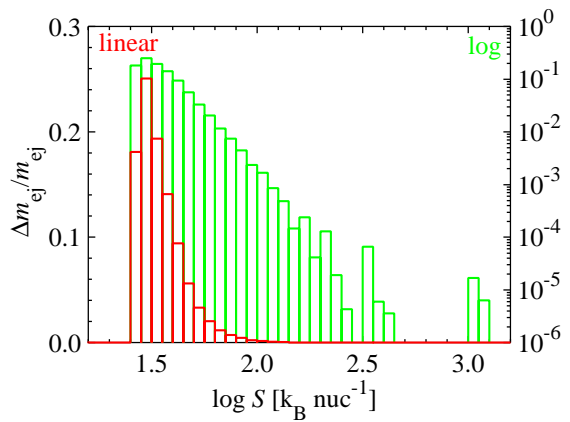}
  \caption{Same as Figure~11, but as a function of the asymptotic value
    of the ejecta entropy $S$.}
\end{figure}

\begin{figure}
  \plotone{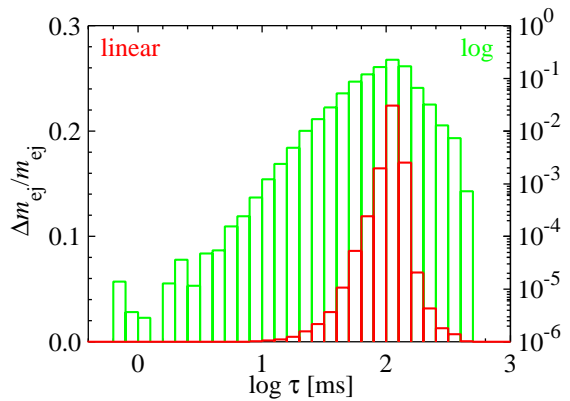}
  \caption{Same as Figure~11, but as a function of the expansion
    timescale $\tau$.}
\end{figure}

From Eq.~(15) and with the numbers of Table~1, the ejecta mass $\Delta
m_\mathrm{ej}$ (normalized by $m_\mathrm{ej}$) as a function of $u$
can be constructed in a binned form as histogram (Fig.~11). We find
that the sub-relativistic winds with $u \sim 0.1 c$ dominate the
ejecta. The relativistic ejecta from the innermost torus are
essentially unimportant. While the entropies are high, $S \sim$
100--1000 $k_\mathrm{B}$, in the innermost winds (Fig.~7), the ejecta
with modest entropies, $S \sim 30 k_\mathrm{B}$, dominate
(Fig.~12). Moreover, the winds with modest $\tau \sim 100$~ms dominate,
and those with short $\tau \sim$ 1--10~ms make a very small
contribution (Fig.~13).

Note that $L_\mathrm{\nu, i}$ as well as $\beta$ in Eqs.~(5) and (6)
would strongly depend on the mass of the torus as well as the viscous
parameter \citep[Fig.~10 in][]{Seti2006}, whose values are highly
uncertain. A higher value of $L_\mathrm{\nu, i}$ will increase the
ejecta mass $m_\mathrm{ej}$ in Eq.~(16) as anticipated from Fig.~5. A
smaller $\beta$ increases the ejecta mass as well (e.g., by about a
factor of two when $\beta = 0.7$, roughly half of the original value,
is chosen) because of a higher $L_{\nu, 0}$ for a longer time (leading
to a higher mass ejection rate; Fig. 5). However, the nucleosynthetic
abundance distributions would not be significantly changed. This is
due to the fact that Eq.~(10) can be written as $Y_\mathrm{e}(t) =
Y_\mathrm{e, f} - (Y_\mathrm{e, f} - Y_\mathrm{e, i})
L_\mathrm{\nu_e}/L_\mathrm{\nu, i}$, i.e., as a function of
$L_\mathrm{\nu_e}$ only (for $\alpha = 1$), being independent of
$\beta$. Because the change of $L_\mathrm{\nu_e}$ is similar to that
of $L_\mathrm{\nu, 0}$ (Table~1), the $Y_\mathrm{e}$ value at a given
$L_\mathrm{\nu, 0}$ is quite similar, e.g., $\Delta Y_\mathrm{e} <
0.02$ for $\beta = 0.7$ is associated with the somewhat shallower
decline of the neutrino luminosities. Accordingly, the integrated
abundance distribution will be mostly independent of $\beta$.

\section{Nucleosynthesis in BH-torus Winds}

The nucleosynthetic yields in each wind trajectory are computed by
solving an extensive nuclear reaction network code. The network
consists of 6300 species between the proton and neutron drip lines,
all the way from single neutrons and protons up to the $Z = 110$
isotopes \citep[for more detail, see][]{Wana2011b}. Neutrino
interactions on free nucleons and $\alpha$-particles are also taken
into account, but fission reactions are not included. As will be
described in \S~4.3, the effect of (neutron-induced) fission during
the $r$-process is expected to be of minor importance in our explored
cases. Our fiducial case (``case~1'') is defined by initiating each
nucleosynthesis calculation when the temperature decreases to $T_9 =
10$ (where $T_9$ is the temperature in units of $10^9$~K), at which
nuclear statistical equilibrium (NSE) is immediately recovered from
arbitrary initial compositions. The initial composition we adopt is
$Y_\mathrm{e}$ and $1 - Y_\mathrm{e}$ for free protons and free
neutrons, respectively, using the $Y_\mathrm{e}$ value of the torus
for a given $L_{\nu, 0}$ (see Table~1). The calculations are carried
out for wind solutions with $L_{\nu, 0} =$ 10--200 in steps of 10 (in
units of $10^{51}$~erg~s$^{-1}$; Table~1) and $D =$ (2--10)
$R_\mathrm{S} =$ 17.7--88.6~km with intervals of $0.1 R_\mathrm{S} =
0.886$~km, i.e., for 1620 trajectories in total.

\subsection{Electron Fractions}

\begin{figure}
  \plotone{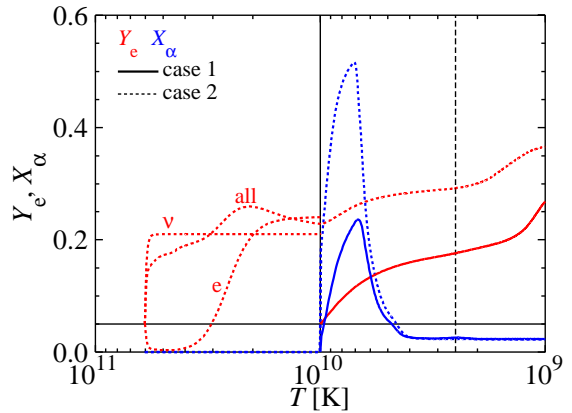}
  \caption{Evolution of the electron fraction ($Y_\mathrm{e}$; red)
    and of the mass fraction of $\alpha$-particles ($X_\alpha$; blue)
    as functions of (decreasing) temperature for case~1 (solid lines;
    only for $T_9 < 10$) and case~2 (dotted lines).  $L_{\nu, 0} = 2
    \times 10^{53}$~erg~s$^{-1}$ and $D = 5 R_\mathrm{S}$ are taken
    for both cases. For case~2, three cases of the $Y_\mathrm{e}$
    evolution are shown (for $T_9 > 10$), in which all four reactions
    (Eqs.~(1)--(4)), only $\nu_\mathrm{e}$ and $\bar{\nu}_\mathrm{e}$
    captures on free nucleons (Eqs.~(3) and (4)), and only $e^-$ and
    $e^+$ captures on free nucleons (Eqs.(1) and (2)) are included
    (indicated by ``all'', ``$\nu$'', and ``e'', respectively). The
    horizontal and vertical solid lines indicate, respectively, the
    initial (torus) $Y_\mathrm{e}$ (= 0.05) and $T_9 = 10$ (at which
    the computation is switched to making use of the full reaction
    network). The vertical dashed line at $T_9 = 2.5$ indicates where
    $r$-processing begins.}
\end{figure}

Figure~14 shows the variation of $Y_\mathrm{e}$ and of the mass
fraction of $\alpha$-particles, $X_\alpha$, as functions of
(decreasing) temperature for $L_{\nu, 0} = 2 \times
10^{53}$~erg~s$^{-1}$ and $D = 5 R_\mathrm{S}$ (case~1; solid
lines). This wind trajectory has the lowest initial (torus)
$Y_\mathrm{e}$ of 0.05 (indicated by a horizontal solid line) of our
considered models. As can be seen in Fig.~14, $Y_\mathrm{e}$ increases
during the early wind expansion phase from its initial value (at $T_9
=10$; vertical solid line) to $Y_\mathrm{e, 2.5}$ (at $T_9 = 2.5$,
considered as corresponding to the onset of the $r$-process phase;
vertical dashed line). This is caused by several effects. The first
effect is that, as soon as the nucleosynthesis calculation starts,
$Y_\mathrm{e}$ relaxes toward the equilibrium value, $Y_\mathrm{e, a}$
\citep[e.g., Eq.~(77) in][]{Qian1996}, due to $\nu_\mathrm{e}$ and
$\bar{\nu}_\mathrm{e}$ captures on free nucleons (Eqs.~(3) and
(4)). The asymptotic value for this wind trajectory is $Y_\mathrm{e,
  a} = 0.21$. The second effect is associated with the continuous
$\alpha$-particle formation (peaking at $T_9 \approx 7$ in Fig.~14)
from recombination of free protons and free neutrons produced by
neutrino capture reactions (Eqs.~(3) and (4)). This drives
$Y_\mathrm{e}$ toward 0.5
\citep[``$\alpha$-effect'';][]{Meye1998a}. As a result of these
combined effects, $Y_\mathrm{e, 2.5}$ (= 0.18) is sizably higher than
the initial value of 0.05.

It is important to note that $\nu_\mathrm{e}$ and
$\bar{\nu}_\mathrm{e}$ captures on free nucleons (Eqs.~(3) and (4))
and also $e^-$ and $e^+$ captures on free nucleons (Eqs.~(1) and (2))
will also operate before the temperature decreases to $T_9 = 10$. We
therefore consider another case (``case~2''), in which each
nucleosynthesis calculation is started from closer to the
neutrinosphere (at $R_\nu = D = 5 R_\mathrm{S}$ and $T_9 = 60$ in this
case) including the four reactions of Eqs.~(1)--(4) (labelled ``all''
in Fig.~14; dotted lines). The computation is then switched to making
use of the full reaction network at $T_9 = 10$. Fig.~14 also shows the
results when including only $\nu_\mathrm{e}$ and
$\bar{\nu}_\mathrm{e}$ captures (``$\nu$''; Eqs.~(3) and (4)) and
$e^-$ and $e^+$ captures (``e''; Eqs.~(1) and (2)) for $T_9 > 10$. We
find that the $Y_\mathrm{e}$ immediately jumps to $\sim 0.15$ mostly
by neutrino captures, continues to gradually increase to 0.26 by
competition of all four reactions of Eqs.~(1)--(4), and decreases
slightly toward $Y_\mathrm{e, a}$ (but still $Y_\mathrm{e} = 0.23 >
Y_\mathrm{e, a} = 0.21$ at $T_9 = 10$). The increase of $Y_\mathrm{e}$
for $T_9 < 10$ is solely due to the $\alpha$-effect. As a result,
$Y_\mathrm{e} = 0.29$ at $T_9 = 2.5$, which is $\sim 0.1$ higher than
that in our fiducial case~1.

\begin{figure}
  \plotone{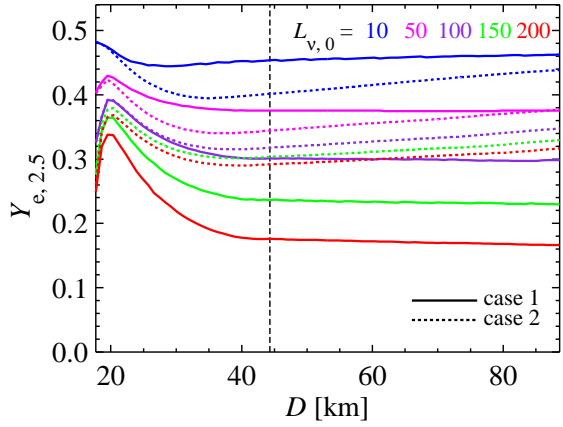}
  \caption{Electron fractions for case~1 (solid lines) and case~2
    (dotted lines) when the temperature has decreased to $2.5\times
    10^9$~K as functions of $D$ for $L_{\nu, 0}$ (in units of
    $10^{51}$~erg~s$^{-1}$) denoted in the legend. The vertical dashed
    line indicates $D = 5 R_\mathrm{S}$.}
\end{figure}

Figure~15 shows the $Y_\mathrm{e}$ values when the temperature has
decreased to $T_9 = 2.5$, $Y_\mathrm{e, 2.5}$, over the entire $D$
range for selected $L_{\nu, 0}$ cases (denoted in the legend). The
shifts of $Y_\mathrm{e}$ from the initial values (see Table~1) are
particularly prominent in the inner winds, where the matter
experiences a strong $\alpha$-rich freezeout owing to the high $S$ and
short $\tau$ (Figs.~7 and 8; cf. Fig.~17). As described above, the
$Y_\mathrm{e, 2.5}$ values are significantly different between cases~1
and 2, in particular for the high $L_{\nu, 0}$ (i.e., low
$Y_\mathrm{e}$) winds.

We emphasize, however, that this is based on our simplified wind
models and the exact $Y_\mathrm{e}$ evolution for $T_9 > 10$ would be
highly dependent on the detailed density structure as well as the
highly uncertain neutrino field in the vicinity of the torus, where
only transport calculations can give an appropriate description of the
energy and direction distribution of the neutrinos. We therefore
consider the $Y_\mathrm{e, 2.5}$ differences between cases~1 and 2 as
possible uncertainties of $Y_\mathrm{e}$ in this study. As can be seen
in Fig.~15, the range of $Y_\mathrm{e, 2.5}$ for case~2 is well
bracketed by that for case~1 between $L_{\nu, 0} = 10$ and 100 (in
units of $10^{51}$~erg~s$^{-1}$). For this reason, we take case~1 as
our fiducial model (and as an optimum case for strong $r$-processing)
in the following sections.

\subsection{Neutron-to-seed Ratios}

\begin{figure}
  \plotone{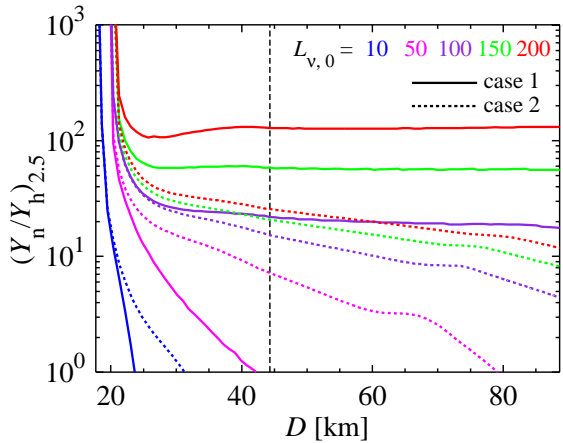}
  \caption{Neutron-to-seed ratios (measured at $T_9 = 2.5$) for case~1
    (solid lines) and case~2 (dotted lines) as functions of $D$ for
    $L_{\nu, 0}$ (in units of $10^{51}$~erg~s$^{-1}$) denoted in the
    legend. The vertical dashed line indicates $D = 5 R_\mathrm{S}$.}
\end{figure}

The neutron-to-seed ratios at the onset of the $r$-process (defined at
$T_9 = 2.5$), $(Y_\mathrm{n}/Y_\mathrm{h})_{2.5}$, are shown in
Figure~16. We find sizable $(Y_\mathrm{n}/Y_\mathrm{h})_{2.5} > 1000$
(up to $\sim 10^8$; not displayed in Fig.~16) in the innermost
region. Except for the inner winds, however,
$(Y_\mathrm{n}/Y_\mathrm{h})_{2.5}$ is nearly constant with $D$ for
$L_{\nu, 0} \ge 100$ (in units of $10^{51}$~erg~s$^{-1}$), being
$(Y_\mathrm{n}/Y_\mathrm{h})_{2.5} \sim 120$ at most. This is large
enough to expect the formation of the third $r$-process peak ($A =
195$) and beyond, but not fission cycling. A high neutron excess
($Y_\mathrm{e, 2.5} \lesssim 0.2$) is needed for a strong $r$-process
at the modest $S$ ($< 100 k_\mathrm{B}$; Figure~7) and $\tau$ ($>
10$~ms; Figure~8) for the winds except for the innermost region.

\subsection{$\alpha$-Particles versus Heavy Nuclei}

\begin{figure}
  \plotone{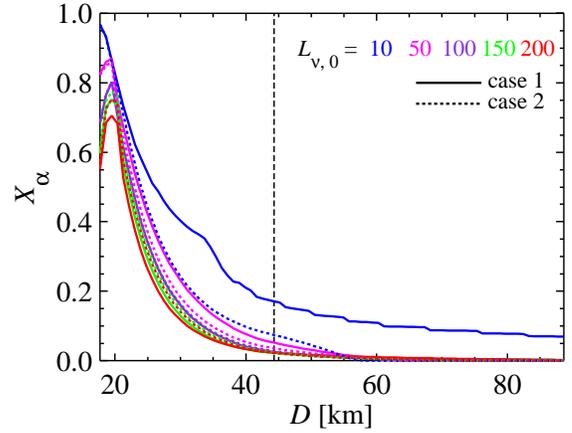}
  \caption{Mass fractions of $\alpha$-particles for case~1 (solid
    lines) and case~2 (dotted lines) as a function of $D$ for $L_{\nu,
      0}$ (in units of $10^{51}$~erg~s$^{-1}$) denoted in the
    legend. The vertical dashed line indicates $D = 5 R_\mathrm{S}$.}
\end{figure}

Figure~17 shows the mass fractions of $\alpha$-particles in the final
nucleosynthetic abundances (solid lines). In the inner winds, the
$\alpha$ concentration is significant owing to the high $S$ and short
$\tau$. The $r$-process thus starts from the seed abundances formed in
neutron-rich quasi nuclear equilibrium \citep[QSE; $A \sim
90$,][]{Meye1998b}. Note that the $\alpha$-abundances do not change
after charged-particle freezeout ($T_9 \sim 4$; Fig.~14). In the
innermost winds, the extremely high $S$ and short $\tau$ lead to
neutron-rich freezeout. This is the reason why $Y_\mathrm{e, 2.5}$
decreases as $D$ approaches the inner tip of the torus (Fig.~15). The
low mass fraction of $\alpha$-particles in the outer winds indicates
that the $r$-process starts from seed abundances formed in conditions
close to neutron-rich NSE \citep[$A \sim 80$,][]{Hart1985,
  Wana2011a}. The nucleosynthetic abundances are thus dominated by
heavy elements, not by $\alpha$-particles, in the outer winds.

\begin{figure}
  \plotone{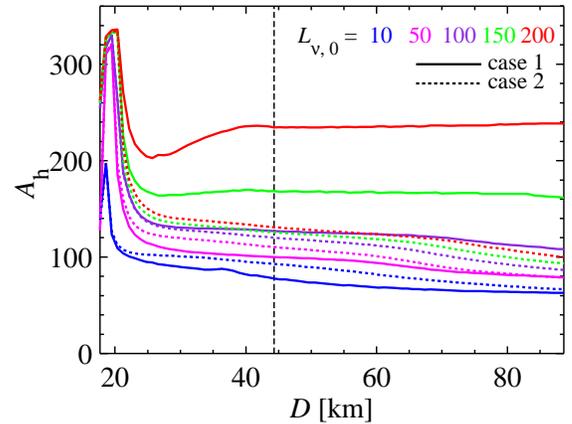}
  \caption{Average atomic mass numbers of heavy nuclei ($Z > 2$) for
    case~1 (solid lines) and case~2 (dotted lines) as functions of $D$
    for $L_{\nu, 0}$ (in units of $10^{51}$~erg~s$^{-1}$) denoted in
    the legend. The vertical dashed line indicates $D = 5
    R_\mathrm{S}$.}
\end{figure}

Figure~18 shows the atomic mass numbers of the final products,
$A_\mathrm{h}$, mass-averaged over heavy nuclei with $Z > 2$. We find
that $A_\mathrm{h}$ is a flat function of $D$ in the outer winds for
all $L_{\nu, 0}$, similar to $(Y_\mathrm{n}/Y_\mathrm{h})_{2.5}$ in
Fig.~16. This suggests that in the outer winds the
torus-$Y_\mathrm{e}$ (assumed to be constant over the entire range of
$D$) predominantly determines $A_\mathrm{h}$, rather than the modest
$S$ or $\tau$, which both exhibit gradients with distance $D$ (Figs.~7
and 8). In the innermost winds ($D < 3 R_\mathrm{S}$), however, the
high $S$ and short $\tau$ play crucial roles. Except for the innermost
winds, $A_\mathrm{h}$ ranges from 60 to 220, encompassing nuclei from
the trans-iron to the actinide region, but well below the
neutron-induced fissioning point \citep[$A \sim 290$, e.g., Fig.~3
in][]{Gori1999}. This is a consequence of $Y_\mathrm{e, 2.5} \ge 0.17$
in the outer winds (Fig.~15), which is still too high to expect
fission cycling at the modest values of $S$ and $\tau$.

\subsection{Mass-integrated Abundances}

\begin{figure}
  \plotone{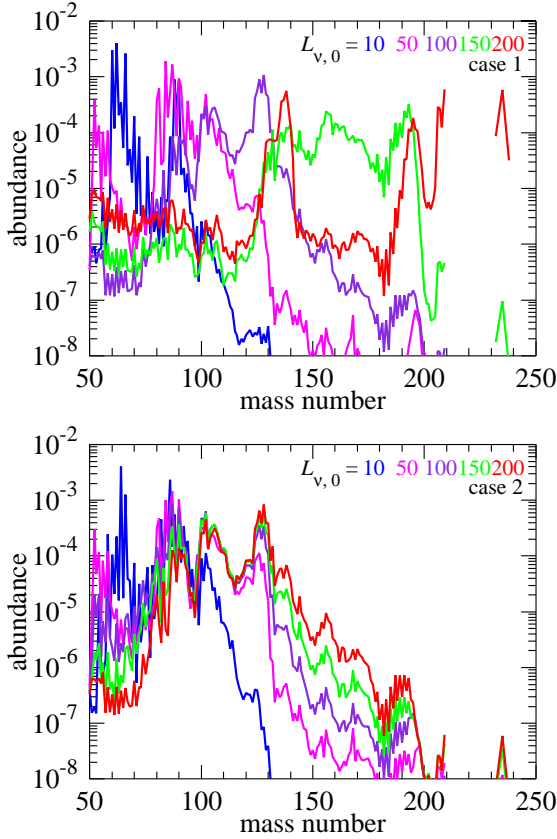}
  \caption{Nucleosynthetic abundances for case~1 (top) and case~2
    (bottom), mass-integrated between $D = 2 R_\mathrm{S}$ and $10
    R_\mathrm{S}$, for the values of $L_{\nu, 0}$ (in units of
    $10^{51}$~erg~s$^{-1}$) denoted in the legend.}
\end{figure}

In order to evaluate the net abundances for each $L_{\nu, 0}$, the
nucleosynthetic yields are mass-integrated over the entire torus range
between $D = 2 R_\mathrm{S}$ and $10 R_\mathrm{S}$. For the $j$-th
$L_{\nu, 0}$, the abundance of nuclide $A$ is calculated with Eqs.~(13)
and (14) as
\begin{eqnarray}
(Y_A)_j
= \frac{1}{\dot{m}_j}
\sum_i (Y_A)_{i, j} \Delta \dot{m}_{i, j}.
\end{eqnarray}
Figure~19 shows the mass-integrated nucleosynthetic abundances for
selected $L_{\nu, 0}$ cases. As noted in \S~4.3, the effect of
neutron-induced fission is expected to be negligible. In order to
roughly include the effect of spontaneous and $\beta$-delayed
fissions, we simply added the abundances with $A \ge 256$ (all
expected to decay by fission) such as
\begin{eqnarray}
Y_{A/2} + 2 Y_A
\longrightarrow 
Y_{A/2} \quad (A \ge 256).
\end{eqnarray}
Actual abundances will depend on the (highly uncertain) decay chains
and the abundance distribution of fission fragments. The sharp
abundance peak at $A \sim 140$ for $L_{\nu, 0} = 2 \times
10^{53}$erg~s$^{-1}$ (case~1; solid red line in Fig.~19) is formed by
fission fragments. The effect of fission for the other cases are
however unimportant.

\begin{figure}
  \plotone{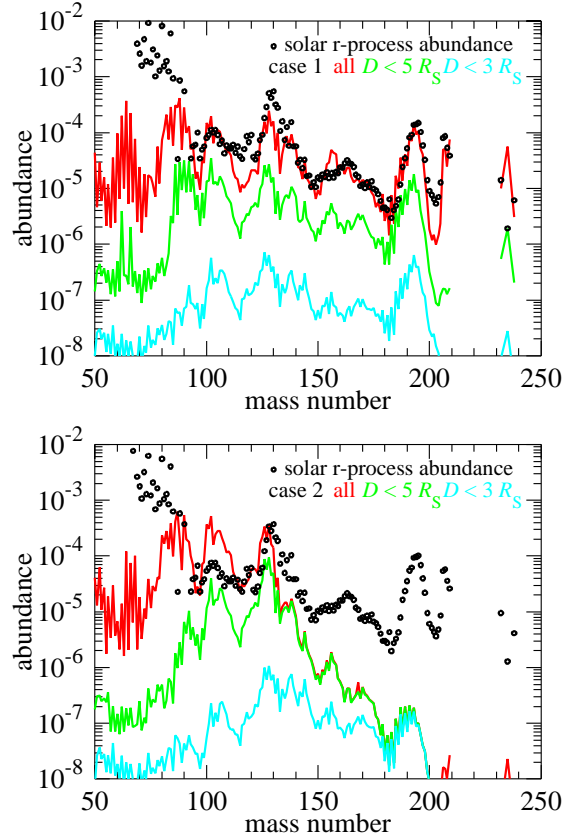}
  \caption{Time-integrated nucleosynthetic abundances for the entire
    torus for case~1 (top panel) and case~2 (bottom panel). The
    calculated abundances for case~1 are in good agreement with the
    solar $r$-process distribution (circles; scaled to match the 3rd
    and 2nd peak heights for case~1 and case~2, respectively). The
    contributions from the inner ($D < 5 R_\mathrm{S}$; green lines)
    as well as innermost ($D < 3 R_\mathrm{S}$; light-blue lines)
    winds are sub-dominant for case~1, but important for case~2.}
\end{figure}

The time-integrated yield of nuclei of atomic mass number $A$ for the
entire torus region is calculated as
\begin{eqnarray}
Y_A
= \frac{1}{m_\mathrm{ej}}
\sum_{i, j} (Y_A)_{i, j} \Delta m_{i, j},
\end{eqnarray}
making use of Eqs.~(15) and (16). In Fig.~20, the resulting yields are
compared with the solar system $r$-process abundances (circles;
vertically shifted to match the height of the third and second
$r$-process peaks for cases~1 and 2, respectively). For case~1, we
find good agreement of the calculated abundances with the solar
$r$-process distribution over the entire range of the $r$-process, $A
= 90-210$. No overproduction of the $N=50$ ($A \approx 90$) nuclei can
be seen, which were problematic in the wind from PNSs born in SNe
\citep{Woos1994, Wana2001}. This can be understood from the fact that
the $r$-processing in the outer winds of BH-tori starts from NSE (or
$\alpha$-deficient QSE) seeds ($A \sim 80$), rather than from
($\alpha$-rich) QSE seeds ($A \sim 90$). For case~2, however, the
production of heavy $r$-process elements beyond $A = 130$ drops
steeply.

Figure~20 also shows the time-integrated abundances from the inner ($D
\le 5 R_\mathrm{S}$) and innermost ($D \le 3 R_\mathrm{S}$) torus. For
case~1, we find similar abundance curves for both regions, but with
sizably smaller amounts than the total production (10 and 100 times
smaller, respectively). This indicates that the low mass ejection
rates (Fig.~5) from the inner and innermost torus diminish the role
of their nucleosynthetic contributions, regardless of their high $S$
and short $\tau$. For case~2, on the other hand, the contributions
from inner and innermost regions dominate the abundances heavier than
$A \sim 140$.

Note that, for case~1, the deficient Pb ($A = 206$--208) and Bi ($A =
209$) abundances relative to the other $r$-elements for $D < 5
R_\mathrm{S}$ (Fig.~20) are due to a fundamental difference in
$r$-processing between the inner and outer winds. The modest $\tau$ of
an outer wind trajectory leads to an $r$-process at high temperature
($T_9 \sim 1$), in which case the nucleosynthetic flow approximately
follows the path determined by the $(n, \gamma)$--$(\gamma, n)$
equilibrium. In the inner region, however, a shorter $\tau$ results in
an $r$-process at substantially lower temperature ($T_9 < 0.5$), in
which the flow is determined by the competition between $(n, \gamma)$
reactions and $\beta$-decays (``cold $r$-process'', Wanajo 2007; also
Blake \& Schramm 1976; Panov \& Janka 2009). \citet{Wana2007} shows
that this non-equilibrium process leads to underabundant Pb.

The ejected mass of $r$-processed nuclei, defined as those with $A \ge
100$, is derived as
\begin{eqnarray}
m_{r, \mathrm{ej}}
= m_\mathrm{ej} \sum_{A \ge 100} A Y_A
= \left \{
\begin{array}{l}
1.30 \times 10^{-3} M_\odot (\mathrm{case~1}),\\
1.16 \times 10^{-3} M_\odot (\mathrm{case~2}),
\end{array}
\right.
\end{eqnarray}
showing a dominance of $r$-processed material (66\% and 59\% for
cases~1 and 2, respectively) in the total ejecta (Eq.~(16)). This is a
consequence of the fact that the $r$-process starts from the NSE (or
$\alpha$-deficient QSE) seeds in most cases (\S~4.3). The mass of
fission products in the ejecta can be calculated as well,
\begin{eqnarray}
m_\mathrm{fis, ej}
= m_\mathrm{ej} \sum_{A \ge 256} A Y_A
= \left \{
\begin{array}{l}
4.25 \times 10^{-5} M_\odot (\mathrm{case~1}),\\
1.19 \times 10^{-7} M_\odot (\mathrm{case~2}),
\end{array}
\right.
\end{eqnarray}
accounting for only 3\% and 0.01\% of the total $r$-process material
for cases~1 and 2, respectively. The mass of Eu in the ejecta, the
element taken as representative of $r$-elements in Galactic chemical
evolution studies \citep[e.g.,][]{Ishi1999}, is estimated to be
\begin{eqnarray}
m_\mathrm{Eu, ej}
= m_\mathrm{ej} \sum_{A = 151, 153} A Y_A
=  \left \{
\begin{array}{l}
6.77 \times 10^{-6} M_\odot (\mathrm{case~1}),\\
3.98 \times 10^{-7} M_\odot (\mathrm{case~2}).
\end{array}
\right.
\end{eqnarray}

\begin{figure}
  \plotone{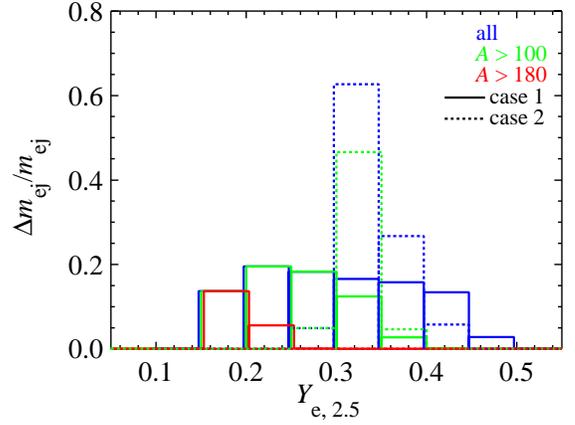}
  \caption{Histogram of the ejecta-mass distribution $\Delta
    m_\mathrm{ej}$ (normalized by the total ejecta mass,
    $m_\mathrm{ej}$) for case~1 (solid lines) and for case~2 (dotted
    lines) as functions of $Y_\mathrm{e, 2.5}$. Different colors
    correspond to all ejecta and those with all yields having $A >
    100$ and those with all yields having $A > 180$, respectively
    (slightly shifted in the horizontal direction for visibility).}
\end{figure}

Figure~21 shows the ejecta-mass histogram as a function of
$Y_\mathrm{e, 2.5}$ for case~1 (solid lines) and case~2 (dotted
lines). The color coding in Fig.~21 discriminates all abundances,
those with all yields having $A > 100$ (all $r$-abundances), and those
with all yields having $A > 180$ (the third peak abundances and
heavier). We find that, for case~1, all $Y_\mathrm{e, 2.5}$ values
contribute with similar weights to the total abundances. For case~2,
the values of $Y_\mathrm{e, 2.5} \approx 0.3$--0.4 dominate the total
ejecta.  Wind trajectories with low $Y_\mathrm{e}$ are, however,
crucial for the production of the $r$-nuclei with $A > 100$
($Y_\mathrm{e, 2.5} \lesssim 0.35$), and in particular with $A > 180$
($Y_\mathrm{e, 2.5} \lesssim 0.20$). This is due to the presence of
modest $S$ and $\tau$ in the ejecta (Figs.~12 and 13), which demands
relatively low $Y_\mathrm{e}$ values for strong
$r$-processing. Figure~21 also shows that the later wind outflow with
$L_\nu < 10^{52}$~erg~s$^{-1}$ ($t \gtrsim 100$~ms and $Y_\mathrm{e}
\sim 0.5$), not included in our calculations, would make no relevant
contribution to the $r$-abundances.

\section{BH-torus Winds as the Origin of \lowercase{$r$}-Elements}

In \S~4, we found that our fiducial model (case~1) of BH-torus winds
leads to the full $r$-process with the solar-like $r$-pattern for $A
=$ 90--210. The ejecta mass of the $r$-processed matter ($A \ge 100$),
$m_{r, \mathrm{ej}} = 1.30 \times 10^{-3} M_\odot$ (Eq.~(20); case~1),
is more than a factor of 10 larger than that needed for CCSNe to be
the dominant source of the $r$-process elements in the Galaxy
\citep{Math1990}. More specifically, the ejected mass of Eu (a nearly
pure $r$-process element), $m_\mathrm{Eu, ej} = 6.77 \times 10^{-6}
M_\odot$ (Eq.~(22); case~1), is a factor of 60 greater than that
needed for CCSNe to be the major source of the Galactic Eu
\citep{Wana2006c}. Provided that our fiducial model (case~1) with
$M_\mathrm{BH} = 3 M_\odot$ and a torus mass of $M_\mathrm{torus} \sim
0.1 M_\odot$ (Eq.~(11)) is representative of NS-NS (or BH-NS) mergers,
the canonical CCSN event rate of $\sim 10^{-2}$~yr$^{-1}$ implies that
a time-averaged Galactic merger rate of $\sim 2 \times
10^{-4}$~yr$^{-1}$ would be needed. This is consistent with the upper
bound of a merger event rate between $7 \times 10^{-6}$ and $3 \times
10^{-4}$~yr$^{-1}$ derived by population synthesis methods for NS-NS
and BH-NS binaries \citep[at solar-metallicity
conditions,][]{Belc2002} and of statistical results based on binary
pulsar surveys \citep[$2 \times 10^{-5}$--$3 \times
10^{-4}$~yr$^{-1}$,][]{Kalo2004}. For case~2 (with inclusion of
reactions of Eqs.~(1)--(4) at $T_9 > 10$), we find inefficient
production of $r$-elements beyond $A \sim 130$. However, a similar
constraint to the Galactic merger rate may be applied for this case,
owing to the similar $r$-processed ejecta mass of $m_{r, \mathrm{ej}}
= 1.16 \times 10^{-3} M_\odot$ (Eq.~(20); case~2).

It is important to note that the actual ejecta mass of $r$-abundances
could be higher than that we obtained on the basis of our simplified
BH-torus wind model. The total ejecta mass, $m_\mathrm{ej} \approx 2
\times 10^{-3} M_\odot$ (Eq.~(16)), can be even 10 times smaller than
that expected from hydrodynamical simulations of compact object
mergers \citep[but not of the subsequent BH-torus accretion with its
neutrino-driven outflow;][]{Jank1999}. There will be additional
contributions from the early mass loss due to the tidal ejection of
neutron-rich matter during the merger event \citep{Frei1999, Gori2005}
and neutrino-driven outflows from HMNSs \citep[for the NS-NS
case,][]{Dess2009} or/and magnetically driven outflows from HMNSs
\citep{Rezz2011, Shib2011}. Moreover, the centrifugal force due to
rapid rotation, which we do not take into account, effectively reduces
the gravity from the central BH. This could lead to larger mass
ejection rates than those obtained here \citep[see][for PNS
winds]{Otsu2000, Wana2001}. Magnetic fields, which are not considered
in our models either, could play a significant role for the mass
ejection, producing viscously driven or MHD-driven outflows in
addition to the considered neutrino-driven ejecta \citep{Metz2009}. It
should also be noted that the production of heavy $r$-process elements
beyond $A \sim 130$ is marginal if we consider our case~2 to be
representative of NS-NS and BH-NS mergers. Case~2 takes into account
$\nu_\mathrm{e}$, $\bar{\nu}_\mathrm{e}$, $e^-$, and $e^+$ captures on
free nucleons also for $T_9 > 10$, resulting in significantly higher
minimal $Y_\mathrm{e, 2.5}$ ($\sim 0.3$) than that in case~1 ($\sim
0.2$). However, the detailed evolution of $Y_\mathrm{e}$ will be
highly dependent on the density structure and the uncertain neutrino
field in the vicinity of the BH-torus, while we consider a simplified
spherical wind model and do not have neutrino transport results for
BH-tori at bound. Effects of magnetic fields noted above would also
modify the evolution of $Y_\mathrm{e}$. General relativistic effects
(which we consider only in the framework of spherically symmetric
configurations) can also be important for determining the exact
$Y_\mathrm{e}$ evolution \citep{Caba2011}. For all these reasons, the
nucleosynthetic outcome of this study should be regarded only as
suggestive. Future detailed multi-dimensional simulations of NS-NS (or
BH-NS) merging, including the later BH-torus wind phase, are required
for more quantitative results.

\begin{figure}
  \plotone{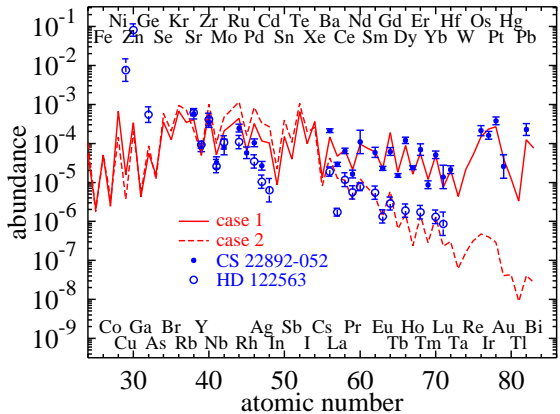}
  \caption{Time-integrated nucleosynthetic abundances for the entire
    torus for case~1 (solid line) and case~2 (dashed line). Cases~1
    and 2 are compared with the spectroscopic abundances of
    $r$-enhanced \citep[CS~31082-001; filled circles,][]{Sned2003} and
    $r$-deficient \citep[HD~122563; open circles,][]{Hond2006}
    Galactic halo stars, respectively. For HD~122563, the Cd and Lu
    values are from \citet{Roed2010} and the Ge value is from
    \citet{Cowa2005}. For both stars, the abundances are vertically
    shifted to match the calculated Eu abundances.}
\end{figure}

Nevertheless, the good agreement of our nucleosynthesis result for
case~1 with the solar $r$-pattern is encouraging. Such an agreement is
also important when one considers compact binary mergers as the origin
of early Galactic $r$-elements in metal-deficient stars with their
uniform (solar-like) $r$-patterns \citep{Sned2008}. In Figure~22, the
nucleosynthetic abundances for case~1 (solid line; as a function of
atomic number) are compared with an $r$-process enhanced star
CS~22892-052 with [Fe/H] $\approx -3.1$ \citep[filled circles; scaled
to match the calculated Eu abundance,][]{Sned2003}. We find quite a
good agreement of our result with the stellar abundances distribution
of CS~22892-052. The good agreement of model and observed abundances
is obtained if our BH-torus wind model represents typical NS-NS (or
BH-NS) merger events, if the minimal $Y_\mathrm{e, 2.5}$ (at the onset
of $r$-processing) is as low as in case~1 ($\lesssim 0.2$), and if the
evolution of $L_{\nu_\mathrm{e}}$ and $L_{\bar{\nu}_\mathrm{e}}$
evolutions in Eqs.~(5) and (6) is appropriate (e.g., $\alpha$ in
Eq.~(6) is not very different from unity and $\beta = 1.3$ and the
factor of 2 within the bracket of Eq.~(6) are good choices). These
should be tested by future long-time hydrodynamical simulations of
NS-NS and BH-NS merging events and of the evolution of their
remnants. Note that, even if the production of $r$-elements heavier
than $A \sim 130$ is marginal as found in case~2, our model could be a
possible explanation for the abundance distribution found in an
$r$-process deficient star HD~122563 with [Fe/H] $\approx -2.7$
\citep[open circles; scaled to match the calculated Eu
abundance,][]{Hond2006}. This suggests that NS-NS and BH-NS mergers
could at least be the origin of some trans-iron elements up to $Z \sim
50$ ($A \sim 120$).

For an observational support, future modeling of the Galactic
$r$-element evolution that is not in conflict with spectroscopic
results of metal-deficient stars will be important
\citep[e.g.][]{Pran2006}. In addition, radioactive decays of
$r$-processed ejecta can lead to faint SN-like transients
\citep{Li1998, Metz2010, Robe2011}. Future detections of such signals,
possibly as accompanying events of short GRBs, will be a direct
support for the occurrence of $r$-processing in NS-NS and BH-NS
mergers. According to the estimates by \citet{Metz2010}, a peak
luminosity of $\sim 3 \times 10^{41}$~erg~s$^{-1}$ could be expected
from the BH-wind ejecta with $m_{r, \mathrm{ej}} \sim 1 \times 10^{-3}
M_\odot$ (Roberts et al. (2011), however, obtained only slightly
higher R-band luminosities for considerably higher ejecta masses). It
will be difficult to distinguish the wind component from the early,
tidally ejected counterpart. Both could contain similar amounts of
radioactive matter. However, the velocities as well as the geometry of
the $r$-processed ejecta might be distinctive between them. Our result
implies that the $r$-processed BH-wind ejecta reach out widely from
the torus with velocities of $\sim 0.1 c$ (except for the innermost
region, Figs.~6 and 11). In contrast, tidal ejecta come from the tips
of (one or two) spiral arms, and are expected to achieve larger
outgoing velocities \citep[$\gtrsim 0.2 c$,][]{Robe2011}. These could
lead to some different features in the light curves (e.g., line
profiles or the time of the peak), potentially distinguishable by
observations.

\section{Summary}

We investigated $r$-process nucleosynthesis in the neutrino-driven
outflows from a BH accretion torus (BH-torus winds) formed in NS-NS
(or BH-NS) mergers. Different from previous works \citep{Surm2008,
  Metz2010} we investigated a time- and space-dependent semi-analytic
model setup. The BH-torus wind models were constructed by considering
spherically symmetric, general relativistic neutrino-driven wind
solutions \citep{Wana2001} with the gravitational potential of a
central BH of $M_\mathrm{BH} = 3 M_\odot$. The BH accretion torus
(around a rotating BH) was assumed to lie between $D = 2 R_\mathrm{S}$
and $10 R_\mathrm{S}$ ($R_\mathrm{S} = 8.86$~km) from the center. Each
wind trajectory reaching away from the torus was obtained by assuming
a hypothetical neutrinosphere in the spherical wind model with the
radius $R_\nu = D$.

In the innermost wind region ($D \sim 2 R_\mathrm{S}$), the efficient
energy deposition due to $\nu \bar{\nu}$ pair annihilation to $e^-
e^+$ pairs leads to very high entropies ($S \gtrsim 100$--$1000
k_\mathrm{B}$ per nucleon) and short expansion timescales ($\tau
\lesssim 1$--10~ms). This allows for a strong $r$-process regardless
of $Y_\mathrm{e}$ \citep[even with $Y_\mathrm{e} > 0.5$ when $\tau <
1$~ms,][]{Meye2002} or no production of heavy elements \citep[in the
relativistic winds with $S \gg 1000 k_\mathrm{B}$,][]{Lemo2002,
  Belo2003}. However, the small mass ejection from the innermost torus
makes this contribution to the total nucleosynthetic abundances
essentially negligible. In the outer wind regions ($D > 5
R_\mathrm{S}$), on the other hand, the dominant heating is due to
$\nu_e$ and $\bar{\nu}_e$ captures on free nucleons as in the case of
PNS winds, resulting in modest entropies ($S \sim 30 k_\mathrm{B}$)
and expansion timescales ($\tau \sim 100$~ms). Low $Y_\mathrm{e}$
values ($\lesssim 0.2$ at $T_9 = 2.5$) are thus essential for a strong
$r$-process. The contribution from outer wind regions dominates the
total nucleosynthetic abundances because of their greater mass
ejection rates. Note that our BH-torus model does not predict a
$\nu$$p$-process \citep{Froe2006, Prue2006, Wana2006a} as suggested in
the case of (proton-rich) collapsar disk winds \citep{Kizi2010}.

The mass-integrated nucleosynthetic abundances, obtained with a
phenomenological time evolution of neutrino luminosities, are in good
agreement with the solar $r$-pattern over the entire $r$-process range
of $A = 90-210$, when the neutrino-matter interactions are considered
only for $T_9 < 10$ (case~1). This can be taken as the optimal case
for strong $r$-processing in our models. However, when
$\nu_\mathrm{e}$, $\bar{\nu}_\mathrm{e}$, $e^-$, and $e^+$ captures on
free nucleons are taken into account also for $T_9 > 10$ (case~2), the
production of heavy $r$-elements beyond $A \sim 130$ drops off
steeply. The total ejecta mass of the BH-torus wind of our simplified
model is calculated to be $m_\mathrm{ej} \approx 2 \times 10^{-3}
M_\odot$, in which the $r$-processed matter dominates ($m_{r,
  \mathrm{ej}} \gtrsim 1 \times 10^{-3} M_\odot$). Provided that our
BH-torus wind model is representative, NS-NS and BH-NS mergers can
produce all (or at least a significant part of) the Galactic
$r$-abundances if the event rate (averaged over the Galactic history)
was $\sim 2 \times 10^{-4}$~yr$^{-1}$, which is consistent with the
upper bound of present (solar-metallicity) population synthesis
results \citep[$7 \times 10^{-6}$--$3 \times
10^{-4}$~yr$^{-1}$,][]{Belc2002} and of statistical results based on
binary pulsar surveys \citep[$2 \times 10^{-5}$--$3 \times
10^{-4}$~yr$^{-1}$,][]{Kalo2004}. This implies that BH-torus winds
from NS-NS and/or BH-NS mergers could be major (or partial) production
sites of the $r$-process elements in the Galaxy. It should be noted
that the actual ejecta mass of $r$-abundances could be substantially
higher than the estimate based on our BH-torus wind model, in which we
do not consider any other effects than neutrino-driven outflows,
namely, centrifugal force, magnetic fields, or tidally ejected
neutron-rich matter from NS disruption.

Obviously, more elaborate hydrodynamical studies of the considered
astrophysical site are needed to obtain information of the neutrino
field that controls the dynamics as well as the neutron-richness in
the BH-torus winds. Note that NS-NS and BH-NS mergers are also
suggested to be the sources of short GRBs. An interesting possibility
in this context is that the radioactive neutron-rich nuclei ejected
during and after mergers might lead to detectable transient
electromagnetic signals \citep{Li1998, Kulk2005, Metz2010,
  Robe2011}. Studies of Galactic chemical evolution will also be
important to test the contributions of NS-NS and BH-NS mergers to the
enrichment history of the $r$-process elements, in particular their
role in the early Galaxy.

\acknowledgements The project was supported by the Deutsche
Forschungsgemeinschaft through the Cluster of Excellence EXC~153
``Origin and Structure of the Universe''
(http://www.universe-cluster.de) and the Transregional Collaborative
Research Centers on ``Neutrinos and Beyond'' (SFB/TR27) and on
``Gravitational-Wave Astronomy'' (SFB/TR7).

\end{document}